\documentclass[12pt]{article}
\usepackage{theorem}
\usepackage{amssymb,amsmath,mathrsfs}
\usepackage{amssymb}
\usepackage{graphicx}
\newtheorem{prop}{}[section]

{\theorembodyfont{\upshape} }
\newcommand{\boma}[1]{{\mbox{\boldmath $#1$} }}

\begin{document}
\def\ellu{\ell_{*}}
\def\dist{\mbox{dist}}
\def\XX{\Xi}
\def\YY{\mbox{{\rm H}}}
\def\xx{\xi}
\def\yy{\eta}
\def\m{m}
\def\cc{\kappa}
\def\Np{\mathfrak N}
\def\Lp{\mathfrak L}
\def\ti{t_1}
\def\Ti{\mathscr T}
\def\Tim{T_1}
\def\kk{K}
\def\K{{\tt K}}
\def\R{{\tt R}}
\def\RR{\mathscr{R}}
\def\Te{\Theta}
\def\ttet{\dot \Theta}
\def\A{{\tt A}}
\def\R{{\tt R}}
\def\L{{\tt L}}
\def\J{{\tt J}}
\def\I{{\tt I}}
\def\X{{\tt X}}
\def\x{{\tt x}}
\def\vel{{\tt v}}
\def\mm{\sigma}
\def\aa{\alpha_{0}}
\def\aad{\alpha_{\delta}}
\def\aadp{\alpha_{\delta'}}
\def\sca{{\scriptstyle{\,\bullet\,}}}
\def\pp{{\mathfrak p}}
\def\qq{{\mathfrak q}}
\def\rr{{\mathfrak r}}
\def\ha{I^1}
\def\ka{I^2}
\def\lau{\lambda_1}
\def\lad{\lambda_2}
\def\tr{\mbox{{\rm tr}}}
\def\v{v}
\def\p{p}
\def\q{q}
\def\u{u}
\def\w{w}
\def\z{z}
\def\Gi{\Lambda}
\def\Gam{\Lambda_{\dag}}
\def\MM{\mathscr{M}}
\def\UU{\mathscr{U}}
\def\FF{\dd{\partial \overline{f} \over \partial I}}
\def\GG{\mathscr{G}}
\def\RR{\mathscr{R}}
\def\NN{\mathscr{N}}
\def\AA{\mathscr{A}}
\def\BB{\mathscr{B}}
\def\CC{\mathscr{C}}
\def\DD{\mathscr{D}}
\def\EE{\mathscr{E}}
\def\LL{\mathscr{L}}
\def\HH{\mathscr{H}}
\def\Ten{\mbox{T}}
\def\Tuu{\Ten^1_1(\reali^{d})}
\def\Tdz{\Ten^2_0(\reali^{d})}
\def\Tud{\Ten^1_2(\reali^d)}
\def\uno{1_d}
\def\ppsi{\vartheta}
\def\PPsi{\Theta}
\def\M{N}
\def\N{V}
\def\half{{1 \over 2}}
\def\II{M}
\def\JJ{N}
\def\ga{\gamma_{\scriptscriptstyle{E}}}
\def\mmu{\nu}
\def\nnu{\mu}
\def\f{g}
\def\Fs{\mathscr{G}}
\def\Kap{\mathscr{K}}
\def\STB{\scriptsize B}
\def\STBB{\scriptsize BB}
\def\STF{\scriptsize F}
\def\STFF{\scriptsize FF}
\def\TB{\scriptsize(B)}
\def\TBB{\scriptsize(BB)}
\def\TF{\scriptsize(F)}
\def\TFF{\scriptsize(FF)}
\def\XXX{\mathscr X}
\def\Piu{\mathscr P}
\def\Men{\mathscr N}
\def\ffi{\varphi}
\def\ES{{\mathcal S}}
\def\KK{{\mathscr K}}
\def\KKp{{\mathscr K}'}
\def\TT{{\mathfrak T}}
\def\kp{k'}
\def\scrscr{\scriptscriptstyle}
\def\scr{\scriptstyle}
\def\dd{\displaystyle}
\def\B{ B_{\mbox{\scriptsize{\textbf{C}}}} }
\def\Bc{ \overline{B}_{\mbox{\scriptsize{\textbf{C}}}} }
\def\ppartial{\overline{\partial}}
\def\d{d}
\def\e{e}
\def\Hinf{ H^{\infty}(\reali^d, \complessi) }
\def\Hn{ H^{n}(\reali^d, \complessi) }
\def\Hm{ H^{m}(\reali^d, \complessi) }
\def\Ha{ H^{\d}(\reali^d, \complessi) }
\def\Ld{L^{2}(\reali^d, \complessi)}
\def\Lpi{L^{p}(\reali^d, \complessi)}
\def\Lq{L^{q}(\reali^d, \complessi)}
\def\Lr{L^{r}(\reali^d, \complessi)}
\def\Knb{K^{best}_n}
\def\D{\mbox{{\tt D}}}
\def\g{ {\textbf g} }
\def\QQQ{ {\textbf Q} }
\def\AAA{ {\textbf A} }
\def\gr{\mbox{graph}~}
\def\Q{$\mbox{Q}_a$~}
\def\PZ{$\mbox{P}^{0}_a$~}
\def\PZAL{$\mbox{P}^{0}_\alpha$~}
\def\PL{$\mbox{P}^{1/2}_a$~}
\def\PU{$\mbox{P}^{1}_a$~}
\def\PK{$\mbox{P}^{k}_a$~}
\def\PKU{$\mbox{P}^{k+1}_a$~}
\def\PI{$\mbox{P}^{i}_a$~}
\def\Pell{$\mbox{P}^{\ell}_a$~}
\def\PTM{$\mbox{P}^{3/2}_a$~}
\def\AZ{$\mbox{A}^{0}_r$~}
\def\AU{$\mbox{A}^{1}$~}
\def\epsilona{\epsilon^{\scriptscriptstyle{<}}}
\def\epsilonb{\epsilon^{\scriptscriptstyle{>}}}
\def\lgraffa{ \mbox{\Large $\{$ } \hskip -0.2cm}
\def\rgraffa{ \mbox{\Large $\}$ } }
\def\restriction{ \stackrel{\setminus}{~}\!\!\!\!|~}
\def\m{m}
\def\Fre{Fr\'echet~}
\def\ap{{\scriptscriptstyle{ap}}}
\def\fiap{\varphi_{\ap}}
\def\BBB{ {\textbf B} }
\def\EEE{ {\textbf E} }
\def\FFF{ {\textbf F} }
\def\TTT{ {\textbf T} }
\def\KKK{ {\textbf K} }
\def\FFi{ {\bf \Phi} }
\def\a{a}
\def\ep{\varepsilon}
\def\parn{\par\noindent}
\def\teta{M}
\def\elle{L}
\def\ro{\rho}
\def\al{\alpha}
\def\si{\sigma}
\def\be{\beta}
\def\de{\delta}
\def\la{{\mathfrak l}}
\def\mi{{\mathfrak v}}
\def\en{{\mathfrak n}}
\def\em{{\mathfrak m}}
\def\te{\vartheta}
\def\tet{\dot \vartheta}
\def\It{\dot I}
\def\Jta{J'}
\def\om{\omega}
\def\ch{\chi}
\def\complessi{{\textbf C}}
\def\reali{{\textbf R}}
\def\interi{{\textbf Z}}
\def\naturali{{\textbf N}}
\def\bT{{\textbf T}}
\def\T1{{\textbf T}^{1}}
\def\Jp{{\hat{J}}}
\def\Pp{{\hat{P}}}
\def\Pip{{\hat{\Pi}}}
\def\Vp{{\hat{V}}}
\def\Ep{{\hat{E}}}
\def\Fp{{\hat{F}}}
\def\Gp{{\hat{G}}}
\def\Ip{{\hat{I}}}
\def\Tp{{\hat{T}}}
\def\Mp{{\hat{M}}}
\def\La{\Lambda}
\def\Si{\Sigma}
\def\Sgr{\Gamma_{\rho}}
\def\DA{\widehat{\Delta}}
\def\DG{\widehat{\Upsilon}}
\def\Lap{{\hat{\Lambda}}}
\def\Sip{{\hat{\Sigma}}}
\def\Upsig{{\check{\Upsilon}}}
\def\Kg{{\check{K}}}
\def\ellp{{\hat{\ell}}}
\def\j{j}
\def\jp{{\hat{j}}}
\def\cir{{\scriptscriptstyle \circ}}
\def\circa{\thickapprox}
\def\vain{\rightarrow}
\def\leqs{\leqslant}
\def\geqs{\geqslant}
\def\ss{s}
\def\vains{\stackrel{\ss}{\rightarrow}}
\def\parn{\par \noindent}
\def\salto{\vskip 0.2truecm \noindent}
\def\spazio{\vskip 0.5truecm \noindent}
\def\vs1{\vskip 1cm \noindent}
\def\fine{\hfill $\diamond$ \vskip 0.2cm \noindent}
\newcommand{\rref}[1]{(\ref{#1})}
\def\beq{\begin{equation}}
\def\feq{\end{equation}}
\def\beqq{\begin{eqnarray}}
\def\feqq{\end{eqnarray}}
\def\barray{\begin{array}}
\def\farray{\end{array}}
\makeatletter \@addtoreset{equation}{section}
\renewcommand{\theequation}{\thesection.\arabic{equation}}
\makeatother
\begin{titlepage}
\begin{center}
{\huge On the average principle \\ for one-frequency systems.}
\end{center}
\vspace{1truecm}
\begin{center}
{\large
Carlo Morosi${}^1$, Livio Pizzocchero${}^2$} \\
\vspace{0.5truecm} ${}^1$ Dipartimento di Matematica, Politecnico
di
Milano, \\ P.za L. da Vinci 32, I-20133 Milano, Italy \\
e--mail: carmor@mate.polimi.it \\
${}^2$ Dipartimento di Matematica, Universit\`a di Milano\\
Via C. Saldini 50, I-20133 Milano, Italy\\
and Istituto Nazionale di Fisica Nucleare, Sezione di Milano, Italy \\
e--mail: livio.pizzocchero@mat.unimi.it
\end{center}
\vspace{1truecm}
\begin{abstract} We consider a perturbed integrable system with one frequency,
and the approximate dynamics for the actions given by averaging over the angle. A classical
qualitative result states that, for a perturbation of order $\ep$, the error of this approximation
is $O(\ep)$ on a time scale $O(1/\ep)$, for $\ep \vain 0$. We replace this with a fully
quantitative estimate; in certain cases, our approach also gives a reliable error estimate
on time scales larger than $1/\ep$. A number of examples are presented; in many cases, our estimator
practically coincides with the envelope of the rapidly oscillating distance between
the actions of the perturbed and of the averaged systems. Fairly good results are
also obtained in some "resonant" cases, where the angular frequency is small
along the trajectory of the system. \parn
Even though our estimates are proved theoretically, their computation in specific
applications typically requires the numerical solution of a system of differential equations.
However, the time scale for this system is smaller by a factor $\ep$ than the time
scale for the perturbed system. For this reason, computation of our estimator is
faster than the direct numerical solution of the perturbed system; the estimator
is rapidly found also in cases when the time scale
makes impossible (within reasonable CPU times) or unreliable the
direct solution of the perturbed system.
\end{abstract}
\vspace{1truecm} \noindent \textbf{Keywords:} Slow and fast motions,
perturbations, averaging method.
\par \vspace{0.4truecm} \noindent \textbf{AMS 2000 Subject
classification}: 70K65, 70K70, 34C29, 70H09, 37J40.
\par
\end{titlepage}
{~}
\vspace{-2cm}
\section{Introduction.}
The averaging method is a classical tool
to analyse dynamical systems with fast
angular variables: the idea is to average over the angles,
to obtain an approximate evolution law for the slow variables (from now on,
called the actions). Many applications are physically relevant; so, error estimates
for this technique on long time scales have an obvious interest.
\parn
Concerning these estimates, the case of one angle is the simplest one
due to the structure of its "resonances", which are produced
only by the vanishing of the angular frequency. However, this one-frequency case covers
non trivial situations: for example, it includes the perturbed Kepler problem,
appearing in applications such as the dynamics of a satellite around an oblate planet and/or
in presence of dragging (see \cite{Ver} and references therein). \parn
The classical theory  for the one-frequency case
states that, under a perturbation $O(\ep)$ of a dynamical system
with one angle and many actions, the difference between the
actions of the perturbed and of the averaged systems is
$O(\ep)$ on a time scale $O(1/\ep)$, for $\ep \vain 0$:
see \cite{Arn} \cite{Arn2} \cite{Bog} \cite{Loc} \cite{Ver} (the
two last references are also useful for general historical and
bibliographical information). This is a qualitative
result; the $n$-th order extensions
of the averaging method proposed in the literature
\cite{Loc} are usually
treated at the same qualitative level, the conclusion being that
some reminder term is $O(\ep^n)$ on a time scale $O(1/\ep)$.
To get these $O(\ep)$ or $O(\ep^n)$ bounds, one generally writes
a number of quite rough majorizations, often containing unspecified constants but
sufficient to obtain a linear integral inequality
for the reminder; the latter is used to obtain the
wanted bounds through the Gronwall Lemma. \parn
Of course, the previously mentioned results are not fully satisfactory if one
aims to obtain precise numerical values from the error analysis; the situation is
especially uncomfortable near resonances, i.e., when the time evolution carries
the system close to a zero of the angular frequency. \parn
In this paper we show that
working carefully, and avoiding unnecessary simplifications,
it is possible to derive fully quantitative and precise error estimates
for the standard ($n=1$) averaging method, for a (small) fixed $\ep$: this requires to solve
a nonlinear integral inequality, or a related differential equation, coupled to a set of auxiliary
differential equations. In typical cases, this is done numerically;
however, the treatment of the above system of equations
is much less expensive than the direct numerical solution of the
action-angle evolution equations; in fact, to get information on an
interval $[0,U/\ep)$ it suffices to solve the previously mentioned
set of equations on the interval $[0,U)$. \parn
To our knowledge, a quantitative error analysis for the averaging
method has been previously proposed in \cite{Smi}; however, in
this reference the attention is mainly focused on specific applications,
admitting a simple analytical treatment, rather than on a general scheme.
In a broader sense, the present paper has some connection with
\cite{MP}; in the cited reference, a quantitative analysis has been proposed
for a rather general class of approximation methods for the evolution equations
(in abstract Banach spaces, so to include the case of evolutionary PDEs).
\salto
\textbf{1A. A precise setting of the problem.}
Let us be given an open set $\Lambda$ of $\reali^d$ (the space of
the actions) and the one-dimensional torus $\bT$:
\beq \Lambda = \{ I = (I^i)_{i=1,...,d} \} \subset \reali^d~, \qquad \bT := \reali/(2 \pi \interi) = \{ \te \}~. \feq
We fix some initial data
\beq I_0 \in \Lambda~, \qquad \te_0 \in \bT~ \label{wef} \feq
and consider
the perturbed one-frequency system
\beq \left\{\barray{ll} d \I/dt = \ep f(\I,\Te)~, &\quad \I(0) = I_0~, \\
d \Te/ d t = \om(\I) +\ep g(\I,\Te)~, &\quad \Te(0) = \te_0 \farray \right. \label{pert} \feq
for two unknown functions $\I : t \mapsto \I(t) \in \Lambda$,
$\Te : t \mapsto \Te(t) \in \bT$. This Cauchy problem contains
the unperturbed frequency
\beq \om \in C^\m(\Gi, \reali)~, \qquad \qquad \mbox{}\om(I) \neq 0 \mbox{ for all $I \in \Gi$}~;
\label{omeg} \feq
the perturbation is governed by a parameter $\ep > 0$, and by two functions
\beq f = (f^i)_{i=1,...,d} \in C^\m(\Gi \times \bT, \reali^d),\qquad
g \in C^\m(\Gi \times \bT, \reali), \label{presc} \feq
$$ (I, \te) \mapsto f(I,\te),~g(I, \te)~; $$
throughout the paper, for
technical reasons it is assumed that $\m \geqs 2$.  \parn
From now
on "the solution $(\I,\Te)$ of \rref{pert}" means the
\textsl{maximal} solution \textsl{in the future}, i.e., the one
with the largest domain of the form $[0,T)$, $T \in (0,+\infty]$
(of course, this domain generally depends on the initial
data). Any expression like "the solution $(\I,\Te)$ exists on $D$"
means that $D$ is a subset of $[0,T)$. It is hardly the case to observe that
$\I,\Te$ are $C^{m+1}$ functions.
\parn The averaged system
associated to \rref{pert} is the Cauchy problem
\beq {d \J \over d \tau} = \overline{f}(\J)~, \qquad  \J(0) = I_0~, \label{av} \feq
$$ \overline{f} = (\overline{f^i}) \in C^\m(\Gi,\reali^d)~, \qquad I \mapsto \overline{f}(I) :=
{1 \over 2 \pi} \int_{\bT} d \te~f(I,\te)~; $$
the unknown is a function $\J : \tau \mapsto \J(\tau) \in \Gi$.
In the same language as before, we stipulate that "the solution $\J$ of \rref{av}" means
the maximal one in the future; again, we have a $C^{m+1}$ function.
\parn
The system \rref{av}
will be compared with \rref{pert} for $\tau = \ep t$, i.e., interpreting $\tau$ as a
rescaled time; if $(\I, \Te)$ is the solution of \rref{pert} and
$\J$ is the solution of \rref{av} \textsl{with the same datum $I_0$ as in} \rref{pert},
the aim is to evaluate the difference $t \mapsto \I(t) - \J(\ep t)$. \parn
The classical result on this subject is an estimate
\beq | \I(t) - \J(\ep \,t) | \leqs C \ep \qquad \mbox{for $t \in [0,1/\ep)$~,} \label{stand} \feq
holding for all sufficiently small $\ep$, under suitable technical conditions
(especially, a lower bound $| \om(I) | \geqs c > 0$ on a convenient domain); in the above,
$C$ is a constant independent of $\ep$. In principle, one could obtain for $C$
a (very complicated) expression, for example
evaluating all the constants in the derivation of \rref{stand} by \cite{Arn};
however, the explicit bound obtained in this way is not satisfactory,
since in typical examples it largely overestimates the difference $\I(t) - \J(\ep t)$~.
\salto
\textbf{1B. Contents of the paper.}
Throughout the paper, the parameter $\ep$ is \textsl{fixed} in $(0,+\infty)$; of course,
our statements are interesting mainly if $\ep$ is small (and are accompanied
by comments which assume this).
Our aim is to perform an accurate analysis of the distance between $\I$ and $\J$; this will ultimately yield a bound
\beq | \I(t) - \J(\ep t) | \leqs \ep \,\en(\ep t) \qquad \mbox{for $t \in [0,U/\ep)$}~, \label{rep} \feq
where $\en : \tau  \mapsto \en(\tau)$ fulfils an integral
inequality, or a related differential equation, for $\tau$ within an interval $[0,U)$. (As we will show,
the existence of $\J,\en$ and some more auxiliary functions for $\tau \in [0,U)$
grants the existence of the solution $(\I,\Te)$ of \rref{pert}
for $t \in [0,U/\ep)$). \parn
Typically, the estimator $\en$ must be computed solving
numerically the above mentioned differential equation; however, this is much less expensive than
the numerical solution of \rref{pert}, because $\en$ depends on the
"slow" time variable $\tau = \ep t$ and thus must be determined
on an interval of length $U$ to get an estimate for $t \in
[0,U/\ep)$ (these considerations can be extended to all the auxiliary functions required in this approach).
In the examples we will present, the function $t
\mapsto \ep \en(\ep t)$ obtained in this way
often coincides with the "envelope" of the rapidly oscillating
function $t \mapsto | \I(t) - \J(\ep t) |$, giving practically the best possible
bound of the form \rref{rep}. Our bound turns out to be fairly good
also in some resonant cases (where $\omega$ vanishes at the boundary of $\Lambda$ and
the actions are close to it, either initially or over long times). As expected, in each example
the CPU time for the computation of $\en$ is much shorter than the CPU time for the
direct solution of \rref{pert}.
\parn
If $U \simeq 1$, Eq. \rref{rep}
can be regarded as a quantitative formulation of the classical
theory, involving the time scales $1$ and $1/\ep$. However, in certain cases
our approach works as well for $U \gg 1$, yielding
accurate estimates for $\! |\I(t) - \J(\ep t)|$ on the extremely large
interval $[0,U/\ep)$; one can even jump to the time scales
$U \simeq 1/\ep$, $U/\ep \simeq 1/\ep^2$.
\parn
The general setting of our approach is described in Section \ref{mainres}, where
we use systematically the function
\beq t \mapsto \L(t) := {1 \over \ep} [\I(t) - \J(\ep t)]~. \feq
After introducing a set of auxiliary functions and differential equations, in Lemma \ref{lemma1} we obtain
an exact integral equation for $\L$; then, in Proposition \ref{mainprop} we derive an integral inequality
and show that any solution $\tau \mapsto \en(\tau)$ of this
inequality gives a bound $| \L(t) | < \en(\ep t)$.
For practical purposes, it is convenient to relate the integral inequality for $\en$ to a
differential equation, which is the subject of Proposition \ref{proprinc};
the solution $\en$ of the differential equation gives a bound $| \L(t) | \leqs \en(\ep t)$,
which is equivalent to Eq.\rref{rep}.
\parn
The subsequent Section \ref{summ} summarizes the path to $\en$, and discusses tests for the
efficiency of this estimator.
The final Section \ref{seces} is devoted to the examples: we mention, in particular, the van
der Pol equation, a resonant case inspired by Arnold, and
Euler's equations for a rigid body under a damping moment linear in the angular velocity
(which also manifest a resonance). \parn
To simplify our exposition, many technical aspects are treated in the Appendices.
In particular: Appendices \ref{aplem1}, \ref{alem} and \ref{apprinc} contain the
proofs of Lemmas \ref{lemma1}, \ref{lemxy} and Proposition \ref{proprinc}, respectively;
Appendices \ref{apol} and \ref{apoll} illustrate the computation of
some auxiliary functions required by the examples of Section \ref{seces}. \parn
The examples presented in this paper are relatively simple, since their purpose is mainly to
test the effectiveness of the method. We postpone to later works (now in progress) the treatment of slightly harder
applications, in particular the already mentioned satellite dynamics.
\salto
\section{Main results.}
\label{mainres}
\textbf{2A. Some notations.}
i)  Vectors of $\reali^d$ are written with upper indices: $X = (X^i)_{i=1,...,d}$~.
We use the spaces $\Ten^p_q(\reali^d)$ of $(p,q)$-tensors over $\reali^d$,
especially for $(p,q) = (1,1), (2,0)$ and $(1,2)$; tensors of these three types are
represented as families of real coefficients
$\AA = (\AA^{i}_{j})$, $\BB = (\BB^{ij})$,
$\CC = (\CC^{i}_{j k})$ ($i,j,k=1,...,d$). \parn
Let $X, Y \in \reali^d$, $\AA, \DD \in \Tuu$, $\BB \in \Tdz$, $\CC \in \Tud$. We define the products
$X Y \in \Tdz$, $\AA X \in \reali^d$, $\AA \DD \in \Tuu$, $\CC X \in \Tuu$, $\CC \BB \in \reali^d$ by
\beq (X Y)^{i j} := X^i Y^j~, \qquad (\AA X)^i := \AA^{i}_{k} X^k~,
\qquad (\AA \DD)^{i}_{j} := \AA^{i}_{k} \DD^{k}_{j}~, \feq
$$ (\CC X)^{i}_{\ell} = \CC^{i}_{k \ell}\, X^k~,\qquad (\CC \BB)^{i} := \CC^{i}_{k \ell}\, \BB^{k \ell} $$
(with the Einstein's summation convention over repeated indices; $X X$ will be written $X^2$).
We note that $\AA \DD$
is the ordinary product of $\AA$ and $\DD$ as matrices;
$\uno$, $\AA^{-1}$ $\in \Tuu$
will denote the identity matrix, and the inverse matrix of $\AA$. The vector $(\CC X) Y = \CC (X Y)$
will be written $\CC X Y$.
\parn
All the considered tensor spaces can be equipped with an inner product $\sca$ and with the corresponding
Euclidean norm $|~|$. If
$X, Y \in \reali^d$, $\AA, \DD \in \Tuu$ and $\CC, \EE \in \Tud$,
\beq X \sca Y := \sum_{i=1}^d X^i Y^i~, \qquad
\AA \sca \DD = \sum_{i, j=1}^d  \AA^{i}_{j} \, \DD^{i}_{j}~,
\qquad
\CC \sca \EE := \sum_{i, j, k=1}^d \CC^{i}_{j k} \,\EE^{i}_{j k}~, \label{sca}
\feq
$$ | X | := \sqrt{X \sca X}~, \qquad
\quad | \AA | := \sqrt{\AA \sca \AA}~, \qquad
| \CC | := \sqrt{\CC \sca \CC}~. $$
\vskip 0.2cm\noindent
ii) Recalling that $\Gi \subset \reali^d$ is open, let $h : \Gi \vain \reali^d$ be $C^\ell$. If
$\ell \geqs 1$ or $\ell \geqs 2$, respectively, the Jacobian and the Hessian
of $h$ at a point $I$ are
\beq {\partial h \over \partial I}(I) := \left({\partial h^i \over \partial I^j}(I)\right)
\in \Tuu~; \qquad
{\partial^2 h \over \partial I^2}(I) := \left({\partial^2 h^i \over \partial I^j \partial I^k}(I)\right)
\in \Tud~. \feq
Let us introduce the set (open in $\reali^d \times \reali^d$).
\beq \Gam := \{ (I,\delta I) \in \Gi \times \reali^d~|~[I, I + \delta I] \in \Lambda \} \label{deg} \feq
(with $[I, I + \delta I]$ denoting the segment of $\reali^d$ with the indicated extremes. For $h$ as before and
$\ell \geqs 1$ or $\ell \geqs 2$, respectively, there are functions $\GG \in C^{\ell-1}(\Gam, \Tuu)$ and
$\HH \in C^{\ell-2}(\Gam, \Tud)$ such that
\beq h(I + \delta I) = h(I) + \GG(I, \delta I) \delta I~, \label{tay0} \feq
\beq h(I + \delta I) = h(I) + {\partial h \over \partial I}(I) \delta I + {1 \over 2} \HH(I,\delta I)
\delta I^2~, \qquad \HH^{i}_{j k}(I, \delta I) = \HH^{i}_{k j}(I, \delta I)~. \label{tay1} \feq
If $d=1$, the above equations can be solved for $\GG$, $\HH$ and determine them
uniquely. If $d>1$, the above equations for $\GG$, $\HH$ have many solutions;
in any dimension, explicit solutions are given by the integral formulas
\beq \GG(I, \delta I) := \int_{0}^{1} d x {\partial h \over \partial I}(I + x \delta I)~,
\quad \HH(I, \delta I) := 2 \int_{0}^{1} d x (1 - x) {\partial^2 h \over \partial I^2}(I + x \delta I)~
\feq
(for $h$ of polynomial or rational type, $\GG$ and $\HH$ can be obtained more directly from
the expression of $h(I + \delta I)$). \parn
In an obvious way, for a function $h : \Gi \times \bT \vain \reali^d$,
we can define the derivatives $({\partial h/\partial I})(I,\te) \in \Tuu$,
$({\partial h/\partial \te})(I,\te) \in \reali^d$,
$({\partial^2 h/\partial I^2})(I,\te) \in \Tud$~.
\vskip 0.2cm\noindent
iii) The average of a $C^\ell$ function $h : \Gi \times \bT \vain \reali^d$
is the $C^\ell$ function $\overline{h} :
\Gi \vain \reali^d$, $I \mapsto \overline{h}(I) := 1/(2 \pi) \int_{\bT} d\te \,h(I,\te)$; this
notation has been already used in Eq. \rref{av}, with $h=f$.
\salto
\textbf{2B. The main Lemma: an integral equation for $\boma{\L}$.}
We consider the perturbed and averaged systems \rref{pert} \rref{av}, for fixed
$\ep > 0$ and initial data $I_0, \te_0$. \parn
The integral equation we are going to derive
will be the basic identity yielding our estimates on $| \L(t) |$; it involves a
number of auxiliary functions, to be introduced as the construction goes on.
\parn
First of all, $s \in C^\m(\Gi \times \bT,\reali^d)$ and
$\p \in C^{\m-1}(\Gi \times \bT, \reali^d)$ are the functions such that
\beq f = \overline{f} + \omega~ {\partial s \over \partial \te}~, \quad
\overline{s} = 0~; \qquad
\p := {\partial s \over \partial I} f + {\partial s \over \partial \te} g~. \label{thef} \feq
The function $s$, which has a preminent role in estimates on $| \L |$, is defined by
\rref{thef} in an implicit way; an explicit formula is
({\footnote{here, $\int_{0}^{\te}$ means integration along any path in $\bT$ from $0$ to $\te$; the integral
depends only on the extremes, because the integrand has zero average. The same could be said for other
integrals appearing later.}})
\beq s = \z - \overline{\z}~, \qquad
\z(I,\te) := {1 \over \om(I)}~\int_{0}^{\te} d \te'~
(f(I,\te') - \overline{f}(I))~. \label{es} \feq
Another function to be used hereafter is
the Jacobian $\FF \in C^{\m-1}(\Gi, \Tuu)$.
From now on, $U$ stands for an element of $(0,+\infty]$.
\begin{prop}
\label{lemma1}
\textbf{Lemma.} Suppose the solution $\J$ of \rref{av}
exists for $\tau \in [0,U)$.
Denote with $\R : [0,U) \vain \Tuu$, $\tau \mapsto \R(\tau)$ and
$\K :  [0,U) \vain \reali^d$, $\tau\mapsto \K(\tau)$
the solutions of
\beq {d \R \over d \tau} = \FF(\J) \,\R~,
\qquad \R(0) = \uno~; \label{sistr} \feq
\beq {d \K \over d \tau} = \FF(\J) \,\K + \overline{\p}(\J)~,
\qquad \K(0) = 0~ \label{sistk} \feq
(these exist and are $C^\m$; $\R(\tau)$ is an invertible matrix for all $\tau \in [0,U)$, and
$\K(\tau) = \R(\tau) \int_{0}^{\tau} d \tau' \, \R(\tau')^{-1} \overline{p}(\J(\tau'))$.
For $d=1$, $\R(\tau) = \exp \int_{0}^{\tau} d \tau' \, \FF(\J(\tau')) \in (0,+\infty)$).
\parn
Furthermore, assume that the solution $(\I,\Te)$ of the perturbed system
\rref{pert} exixts for $t \in [0, U / \ep)$, with $(\J(\ep t), \I(t) - \J(\ep t)) \in \Gam$. Finally, define
\beq \L : [0,U/\ep) \vain \reali^d~, \qquad
t \mapsto \L(t) :=
{1 \over \ep} \,[ \I(t) - \J(\ep t) ]~. \label{defr} \feq
Then, for $t \in [0,U/\ep)$ it is
\beq  \L(t) = s(\I(t), \Te(t)) - \R(\ep t) \, s(I_0, \te_0) - \K(\ep t) + \label{inseq}\feq
$$ - \ep \Big(\w(\I(t), \Te(t)) - \FF(\J(\ep t)) \, \v(\I(t), \Te(t)) \Big)~+  $$
$$ + \,\ep^2 \R(\ep t) \int_{0}^t d t' \,
\R^{-1}(\ep t') \Big(\u(\I(t'),\Te(t')) -\FF(\J(\ep t')) (\w +\q)(\I(t'),\Te(t')) +  $$
$$ - \MM(\J(\ep t')) \v(\I(t'),\Te(t')) - \GG(\J(\ep t'), \ep  \L(t')) \L(t')
+ {1 \over 2} \HH(\J(\ep t'), \ep \L(t')) \, \L(t')^2\Big)~. $$
In the above, $v \in C^\m(\Gi \times \bT,\reali^d)$,
$q, \w \in C^{\m-1}(\Gi \times \bT, \reali^d)$,
$\u \in C^{\m-2}(\Gi \times \bT, \reali^d)$,
and $\MM \in C^{\m-2}(\Gi, \Tuu)$ are the functions uniquely defined by the following equations:
\beq s = \omega {\partial \v \over \partial \te}~, \qquad
\v(I,\te_0) = 0 \quad \mbox{for all $I \in \Gi$;} \label{st} \feq
\beq \q := {\partial \v \over \partial I} f + {\partial \v \over \partial \te} g~;
\label{dq}  \feq
\beq \p = \overline{\p} + \omega {\partial \w \over \partial \te}~, \qquad
\w(I,\te_0) = 0 \quad \mbox{for all $I \in \Gi$}~; \label{qw} \feq
\beq \u := {\partial \w \over \partial I} f + {\partial \w \over \partial \te} g~; \qquad
\MM := {\partial^2 \overline{f} \over \partial I^2}\, \overline{f}  - \left(\FF\right)^2 ~. \label{em} \feq
Furthermore, $\GG \in C^{\m-2}(\Gam, \Tuu)$ and $\HH
\in C^{\m-2}(\Gam,\Tud)$ are two functions fulfilling Eq.s \rref{tay0} for
$h=\overline{p}$ and \rref{tay1} for $h=\overline{f}$: so,
for $(I, \delta I) \in \Gam$,
\beq \overline{p}(I + \delta I) = \overline{p}(I) + \GG(I, \delta I) \delta I~, \label{equel0} \feq
\beq \overline{f}(I + \delta I) = \overline{f}(I) + \FF(I) \delta I
+ {1 \over 2} \HH(I,\delta I) \delta I^2~,
\quad \HH^{i}_{j k}(I, \delta I) = \HH^{i}_{k j}(I, \delta I)~. \label{equel} \feq
\end{prop}
\textbf{Proof.} It is obtained by a long computation, where the functions
$s,...., \HH$ appear gradually. See the Appendix \ref{aplem1}. \fine
\textbf{Remarks. i)} The above definitions must be understood in terms of the previous tensor notations;
for example, the
equivalent formulation in components of Eq. \rref{em} is
$\dd{\MM^{i}_{j} = {\partial^2 \overline{f^i} \over \partial I^k \partial I^j}\, \overline{f^k} -
{\partial \overline{f^i} \over \partial I^k} \, {\partial \overline{f^k} \over \partial I^j}}$.
Of course
$\v(I,\te) = \om^{-1}(I) \int_{\te_0}^{\te} d \te'~s(I,\te')$,
$\w(I,\te) = \om^{-1}(I)) \int_{0}^{\te} d \te'~
(\p(I,\te') - \overline{\p}(I))$. \parn
\textbf{ii)} If we write $\I(t) = \J(\ep t) + \ep \L(t)$, \rref{inseq} becomes
an integral equation for $\L$. Most of the terms therein are slow, i.e.,
depend on $\ep t$: the exceptions are $\L$ itself and the angle $\Theta$.
The subsequent step after this Lemma will be to infer from \rref{inseq}
an integral inequality involving only the slow time variable $\ep t$; we note that,
even though the integral in Eq. \rref{inseq} is multiplied by $\ep^2$, this term
appears to be of order $\ep$ if we consider $\ep t'$ as the integration variable.
In any case, the presence of a small factor $\ep$ in front of the integral allows us
to use for it fairly rough estimates.
\salto
\textbf{2C. A second Lemma: an integral inequality for $|\L|$.}
Throughout this paragraph we assume that the solution $\J$ of the averaged system exists on $[0,U)$, and
define $\R,\K$ via Eq.s \rref{sistr}, \rref{sistk}.
$B(I,\varrho)$ denotes the
open ball in $\reali^d$ of center $I$ and radius $\varrho$;
we furtherly suppose the following. \parn
i) There is a function $\ro \in C([0,U), [0,+\infty])$ such that
\beq B(\J(\tau), \rho(\tau)) \subset \Lambda \qquad \mbox{for $\tau \in [0,U)$}~. \label{recall} \feq
We denote with $\Sgr$ the subgraph of $\rho$, i.e.,
\beq \Sgr := \{ (\tau, r)~|~\tau \in [0,U),~r \in [0,\ro(\tau))~ \}~. \label{dero} \feq
ii) There are functions
\beq a,b,c,\d,\e \in C(\Sgr, [0,+\infty))~\feq
such that, for any $\tau \in [0,U)$, $\delta J
\in B(0, \rho(\tau))$ and $\te \in \bT$,
\beq | s(\J(\tau)+\delta J,\te) - \R(\tau) s(I_0,\te_0) - \K(\tau) | \leqs a(\tau,|\delta J|)~, \label{fa} \feq
\beq \left| \w(\J(\tau) +\delta J, \te) - \FF(\J(\tau)) \, \v(\J(\tau) + \delta J, \te) \right|~
\leqs b(\tau, |\delta J|)~, \label{fb} \feq
$$ \Big| \u(\J(\tau)+\delta J, \te) -\FF(\J(\tau)) (\w +\q)(\J(\tau)+\delta J,\te) + $$
\beq - \MM(\J(\tau)) \v(\J(\tau)+\delta J,\te) \Big|
\leqs c(\tau,|\delta J|)~, \label{fc} \feq
\beq | \GG(\J(\tau),\delta J)| \leqs \d(\tau,|\delta J|)~, \label{fd} \feq
\beq | \HH(\J(\tau),\delta J)| \leqs \e(\tau,|\delta J|)~ \label{fe} \feq
(note that $(\J(\tau), \delta J)
\in \Gam$, by the convexity of the sphere).
The functions $c, d, e$ are assumed to be non decreasing with respect to the second variable:
\beq (\tau,r), (\tau, r') \in \Sgr,~~ r \leqs r'~~~\Rightarrow~~~ c(\tau,r) \leqs c(\tau,r')~,~~ \label{monot} \feq
and similarly for $d, e$.
Given $a, b,c, \d,\e$, we define the functions
\beq \alpha \in C(\Sgr, [0,+\infty)), \quad
\alpha(\tau,r) := a(\tau,r) + \ep b(\tau,r)~,\label{al} \feq
\beq \gamma \in C(\Sgr \times [0,+\infty), [0,+\infty)), \quad
\gamma(\tau,r,\ell) := c(\tau, r) + \d(\tau, r) \ell + {1 \over 2} \e(\tau, r) \ell^2~. \label{ga} \feq
We can now write the integral inequality for the function $t \mapsto | \L(t) |$, with $\L$ as in \rref{defr}.
\begin{prop}
\label{seclem}
\textbf{Lemma} Assume that the solution $(\I,\Te)$ of
the perturbed system exists on $[0,U/\ep)$, and that $|\L(t)| < \ro(\ep t)/\ep$
for all $t \in [0,U/\ep)$. Then
\beq | \L(t) |  \leqs \alpha(\ep t,\ep | \L(t) |)
+ \ep^2 | \R(\ep t) | \int_{0}^t d t' \, | \R^{-1}(\ep t') |\, \gamma(\ep t', \ep | \L(t') |,
| \L(t') |)~. \label{inseqp} \feq
\end{prop}
\textbf{Proof.} We take the norm of both sides in Eq. \rref{inseq}.
To estimate the right hand side, we use some Schwarz inequalities and Eq.s (\ref{fa}--\ref{fe}) with
$\delta J = \I(t) - \J(\ep t) = \ep \L(t)$; then, the thesis follows from
the definitions \rref{al} \rref{ga} of $\alpha$ and $\gamma$. \fine
\salto
\textbf{2D. A third Lemma, on integral inequalities.} To go on, we need a
general result on a class of integral inequalities; we state it at an abstract level,
forgetting momentarily the function $|\L|$.
\begin{prop}
\label{lemxy}
\textbf{Lemma.} Let $T \in (0,+\infty]$, $\delta \in C([0,T), [0,+\infty])$ and
\beq \XX := \{ (t, \ell)~|~ t \in [0,T), \ell \in [0,\delta(t))~\}, \feq
$$ \YY := \{ (t, t', \ell)~|~t \in [0,T)~, t' \in [0,t],~ (t', \ell) \in \XX~\}~. $$
Consider two functions $\xx \in C(\XX, [0,+\infty))$ and
$\yy \in C(\YY, [0,+\infty))$, the latter non decreasing
in the last variable: $\yy(t,t', \ell') \leqs \yy(t,t', \ell)$ for $(t,t',\ell) \in \YY$ and
$\ell' \in [0,\ell]$. Furthermore,
let $\la \in C([0,T), [0,+\infty))$ and
$\mi \in C([0,T), (0,+\infty))$ be such that $\mbox{graph}~ \la$, $\mbox{graph}~ \mi$
$\subset \XX$, and
\beq \la(0) = 0~, \qquad
\la(t) \leqs \xx(t, \la(t)) + \int_{0}^t d t' \, \yy(t, t', \la(t'))~, \label{ero} \feq
\beq \mi(t) > \xx(t, \mi(t)) + \int_{0}^t d t' \, \yy(t, t', \mi(t'))~ \label{esi} \feq
for all $t \in [0,T)$. Then
\beq \la(t) < \mi(t) \qquad \mbox{for all $t \in [0,T)$}~.\label{ete} \feq
\end{prop}
\textbf{Proof.} It adapts the one of a similar result in \cite{Mitr}; see the Appendix \ref{alem}. \fine
\salto
\textbf{2E. The main Proposition.}
Throughout this paragraph we still assume that the solution $\J$ of the averaged system
exists on $[0,U)$, and
define $\R,\K$ via Eq.s \rref{sistr} \rref{sistk}. We also assume
there is a set of functions $\rho,a,b,c,d,e$ as in paragraph 2C; $\alpha$ and
$\gamma$ are defined consequently, as indicated therein.
\begin{prop}
\label{mainprop}
\textbf{Proposition.} Assume that there is a
function $\en \in C([0,U),(0,+\infty))$ such that, for all $\tau \in [0,U)$,
\beq \en(\tau) < \ro(\tau)/\ep~, \label{rc} \feq
\beq \en(\tau) >
\alpha(\tau,\ep \en(\tau))
+ \ep | \R(\tau) | \int_{0}^{\tau} d \tau' \, | \R^{-1}(\tau') | \,  \gamma(\tau', \ep \en(\tau'), \en(\tau'))~.
\label{inecont} \feq
Then, the solution $(\I,\Te)$ of the perturbed system exists on
$[0,U/\ep)$; furthermore, defining $\L$ as in Eq. \rref{defr} we have
\beq | \L(t) | < \en(\ep t) \qquad \mbox{for all $t \in [0,U/\ep)$.} \feq
\end{prop}
\textbf{Proof.} Let us recall that $(\I,\Te)$ is the
\textsl{maximal} solution of \rref{pert}, and denote its domain with $[0, V/\ep)$; for the moment,
this merely defines the coefficient $V \in (0,+\infty]$ (which can
depend on $\ep$ and be large, small, etc.). To go on,
we provisionally put
\beq U' := \min(V, U)~; \feq
one of our aims is to show that $U' = U$, but this will be established
only in the second step of the proof. We also define $\L$ as in Eq.
\rref{defr}, but on the domain $[0,U'/\ep)$. \parn
\textsl{Step 1.  One has}
\beq | \L(t) | < \en(\ep t) \qquad \mbox{for all $t \in [0,U'/\ep)$}~.
\label{ijs} \feq
To show this, we write the integral inequality \rref{inecont} with $\tau= \ep t$, $\tau' = \ep t'$; this gives
\beq \en(\ep t) >
\alpha(\ep t,\ep \en(\ep t))
+ \ep^2 | \R(\ep t) | \int_{0}^{t} d t' \, | \R^{-1}(\ep t') |
\, \gamma(\ep t', \ep \en(\ep t'), \en(\ep t'))
\label{ineconttt} \feq
for all $t \in [0,U/\ep)$, and \textsl{a fortiori} for $t \in [0,U'/\ep)$. \parn
On the other hand, Lemma \ref{seclem} can be applied with the constant $U$ therein replaced by $U'$,
because $(\I, \Te)$ is defined on $[0,U'/\ep)$ and $\J$ is defined on $[0,U')$; thus,
Eq. \rref{inseqp} for $|\L(t)|$ holds for $t \in [0,U'/\ep)$.
Now, we apply Lemma \ref{lemxy} with
\beq T := {U' \over \ep}~, \qquad  \delta(t) := \ro(\ep t)/\ep, \feq
$$ \xx(t,\ell) :=
\alpha(\ep t,\ep \ell)~, \quad
\yy(t, t', \ell) := \ep^2\, | \R(\ep t) |\, | \R^{-1}(\ep t') | \, \gamma(\ep t', \ep \ell, \ell)~, $$
$$ \la(t) := |\L(t)|~, \qquad \mi(t) := \en(\ep t)~; $$
of course, the initial condition $\la(0) = 0$ holds because $\I(0) = I_0 = \J(0)$.
Lemma \ref{lemxy} gives $\la(t) < \mi(t)$, which is just the relation  \rref{ijs}. \parn
\textsl{Step 2. It is}
\beq U' = U \feq
(\textsl{thus $(\I, \Te)$ exists on $[0,U/\ep)$, and the inequality of
Step 1 holds on this interval)}. \parn It suffices to show that $V \geqs U$; to this purpose we suppose $V < U$,
and infer a contradiction. Indeed, let us put
\beq K := \{ (t, I) \in [0,V/\ep] \times \reali^d~|~| I - \J(\ep t)| \leqs \ep \en(\ep t) \}~. \feq
This is a closed subset of $\reali \times \reali^d$; it is bounded, since $t \mapsto \J(\ep t)$,
$t \mapsto \en(\ep t)$ are bounded functions on $[0, V/\ep]$. Thus, $K$ is a compact subset
of $\reali \times \reali^d$. We note
that $(t,I) \in K$ implies $I \in \overline{B}(\J(\ep t), \ep \en(\ep t)) \subset B(\J(\ep t), \ro(\ep t))
\subset \Lambda$ (recall Eq.s \rref{rc} and \rref{recall}); thus,
$K \subset [0,V/\ep] \times \Lambda$. \parn
The previous considerations ensure compactness of $K \times \bT$ $\subset$ $\reali \times \Lambda \times \bT$;
due to Step 1, we have $\mbox{graph}~(\I, \Te) \subset K \times \bT$~.
The inclusion into a
compact set and a standard continuation principle  for ordinary differential equations \cite{Zei} imply
that the solution $(\I, \Te)$ can be extended to an interval larger than $[0,V/\ep)$. This contradicts
our maximality assumption, and concludes the proof. \fine
\salto
\textbf{2F. A differential reformulation of the previous results.}
For practical applications, and especially for the numerical implementation of our scheme by standard
packages, it is convenient to replace the integral inequality \rref{inecont} for $\en$ with a
differential equation related to it. This
equation is presented hereafter, and will be the basis of all applications discussed
in the next sections; it is supplemented by an initial condition, defined implicitly by a fixed point problem.
\parn
In the sequel we keep the assumptions at the beginning of paragraph 2E, but we require some more
regularity on the functions
$a,b,c,d,e$ fulfilling Eq.s (\ref{fa}--\ref{fe}), namely,
\beq a,b \in C^2(\Sgr,\reali)~, \qquad
c, \d, \e \in C^1(\Sgr, \reali)~; \label{assug} \feq
so, the functions $\alpha, \gamma$ in Eq.s \rref{al} \rref{ga} are, respectively,
of class $C^2$ and $C^1$.
\begin{prop}
\textbf{Proposition.}
\label{proprinc}
i) Assume there are real numbers $\ellu, M \geqs 0$ and $\mm > 0$ such that
\beq \Sigma := [\ellu - \mm, \ellu + \mm] \subset (0,\ro(0)/\ep)~, \feq
\beq M < 1/ \ep~, \qquad \left|{\partial \alpha \over \partial r}(0, \ep \ell) \right| \leqs M
\quad \mbox{for $\ell \in \Sigma$}~, \label{hi} \feq
\beq | \alpha(0, \ep \ellu) - \ellu | + \ep M \mm < \mm~. \label{hip} \feq
Then, the map $\ell \mapsto \alpha(0, \ep \ell)$ sends the interval $\Sigma$ into itself
and is therein contractive with Lipschitz constant $\ep M$. So, there is a unique
$\ell_0 \in \Sigma$ solving the fixed point equation
\beq \alpha(0, \ep \ell_0) = \ell_0~. \label{fixedp} \feq
ii) With $\ell_0$ as above, let
$\em, \en \in C^1([0,U),\reali)$ solve the Cauchy problem
\beq {d \em \over d \tau} =
| \R^{-1} | \, \gamma(\cdot , \ep \en, \en)~, \qquad
\em(0) = 0~, \label{tre} \feq
$$ {d \en \over d \tau} = \Big(1 - \ep {\partial \alpha \over \partial r}\, (\cdot, \ep \en)\Big)^{-1}
\left( {\partial \alpha \over \partial \tau}\,(\cdot , \ep \en)
+ \ep | \R | | \R^{-1} | \, \gamma(\cdot , \ep \en, \en) + \ep | \R |^{-1}~\big(\R \sca {d \R \over d \tau}
\big)~\em \right)~, $$
\beq \qquad \en(0) = \ell_0~, \label{quattro} \feq
with the domain conditions
\beq 0 < \en < \ro/\ep~, \qquad
{\partial \alpha \over \partial r}\,(\cdot, \ep \en)  < 1/\ep  \label{due} \feq
(note that \rref{tre} implies
$\em \geqs 0$; in the above, $\sca$ is the inner product of Eq.
\rref{sca}). \parn
Then, the solution $(\I,\Te)$ of the perturbed system exists on
$[0,U/\ep)$ and (with $\L$ as in \rref{defr})
\beq | \L(t) | \leqs \en(\ep t) \qquad \mbox{for all $t \in [0,U/\ep)$.} \feq
\end{prop}
\textbf{Proof.} It is found in the Appendix \ref{apprinc}, after a necessary lemma. \fine
\vskip 0.2cm\noindent
\section{A summary of the method, and how to test it.}
\label{summ}
\textbf{3A. The main steps to implement the scheme of the previous section.}
In the approach we have outlined, the steps to be performed are the following ones. \parn
i) Compute
$\overline{f}$ and the functions $s, p, ..., \MM, \GG, \HH$ of Eq.s \rref{thef}  (\ref{st}--\ref{equel}). \parn
ii) Determine the solution
$\J$ of Eq. \rref{av}, on some interval $[0,U)$; solve
Eq.s \rref{sistr} \rref{sistk} for $\R, \K$ on the same interval. \parn
iii) Find a set of functions $\rho,a,b,c,d,e$ as in paragraph 2C, so as to
fulfil the inequalities (\ref{fa}--\ref{fe}); from them, define the functions
$\alpha, \gamma$ via Eq.s \rref{al} \rref{ga}. In the subsequent steps, we make on
$a,...,e$ the assumptions \rref{assug}. \parn
iv) Determine $\ell_0$, solving the fixed point problem \rref{fixedp}. \parn
v) Search for functions $\em,\en$ fulfillfing Eq.s \rref{tre} \rref{quattro},
with the domain conditions \rref{due}. If these equations and \rref{av}
have solutions on some interval $[0,U)$, we can grant existence
on $[0,U/\ep)$ for the solution
$(\I,\Te)$ of \rref{pert}, and we know that $\L(t)  := (\I(t) - \J(\ep t))/\ep$
fulfils on this interval the bound $| \L(t) | \leqs \en(\ep t)$.
\vskip 0.2cm \noindent
Here are some general comments on the practical implementation of the previous steps
(these will also be useful to introduce the examples of the next section). \parn
i) Of course, the computation of $\overline{f},s, p, ..., \MM$ is more or less difficult
depending on $f$, $g$ and $\omega$, concerning especially the integrals over $\te$. These
computations can involve special functions (it should be noted that, in many
examples coming from mechanics,
$f$, $g$ and $\omega$ are themselves special functions). Generally, the determination of
$\overline{f},s, p, ..., \MM$ is simple when, for fixed $I$, $f$ and $g$ are trigonometric
polynomials in $\te$. Concerning $\GG$ and $\HH$, see the remarks that
follow Eq.s \rref{tay0} \rref{tay1}. \parn
ii) The determination of $\J,\R,\K$ will be analytical in the symplest cases, and
otherwise numerical.
\parn
iii) For the implementation of our scheme, the functions
$b,c,d,e$ are slightly less important than $a$; in fact,
they are always multiplied by the small parameter $\ep$
whenever they appear in steps iii) iv) v). For this reason, it is important
to compute $a$ estimating as accurately as possible the left hand side in Eq. \rref{fa}; as for
$b,...,e$, in many cases one can accept rougher majorizations for the left hand sides of Eq.s
(\ref{fb}--\ref{fe}). \parn
In many applications, such as in the examples of the next section, the functions $a,b,..,e$ will have the form
\beq a(\tau,r) := \widehat{a}(\J(\tau), \R(\tau), \K(\tau), r),~
b(\tau, r) := \widehat{b}(\J(\tau), r), ...,~ e(\tau, r) = \widehat{e}(\J(\tau), r) \label{defby} \feq
depending on certain known functions
\beq \widehat{a} \in C^2(\DA, \reali)~, \quad \widehat{b}
\in C^2(\DG, \reali)~, \quad
\widehat{c},\widehat{d},\widehat{e} \in C^1(\DG, \reali)~, \feq
with domains
\beq \DA \subset \reali^d \times \Tuu \times \reali^d \times \reali~\mbox{open},~~~
\DG \subset \reali^d \times \reali~\mbox{open ~ such that} \feq
$$ (\J(\tau), \R(\tau), \K(\tau), r)
\in \DA~, \qquad (\J(\tau), r) \in \DG
\quad \mbox{for all} \quad (\tau,r) \in \Sgr~. $$
Of course, in this case it is
\beq \alpha(\tau,r) = \widehat{\alpha}(\J(\tau), \R(\tau), \K(\tau), r)~, \quad
\gamma(\tau, r, \ell) = \widehat{\gamma}(\J(\tau), r, \ell)~, \label{casealp} \feq
where $\widehat{\alpha} \in C^2(\DA, \reali)$ and $\widehat{\gamma} \in
C^1(\DG \times \reali, \reali)$
are defined by
\beq \widehat{\alpha}(J,\RR,\kk,r) := \widehat{a}(J,\RR, \kk, r) + \ep \widehat{b}(J,r)~,\label{aal} \feq
\beq \widehat{\gamma}(J,r,\ell) := \widehat{c}(J, r) + \widehat{d}(J, r) \ell + {1 \over 2} \widehat{e}(J, r) \ell^2~. \label{gga} \feq
Furthermore, the derivative $\partial \alpha/\partial \tau$ in Eq. \rref{quattro} is given by
\beq {\partial \alpha \over \partial \tau}(\cdot, r) =
{\partial \widehat{\alpha} \over \partial J}\,(\J,\R,\K,r) \sca {d \J \over d \tau}+
{\partial \widehat{\alpha} \over \partial \RR}\,(\J, \R,\K, r) \sca {d \R \over d \tau}
+ {\partial \widehat{\alpha} \over \partial \kk} \,(\J, \R,\K, r) \sca {d \K \over d \tau}
\label{eqquattro} \feq
with $\partial \widehat{\alpha}/\partial \RR :=
(\partial \widehat{\alpha}/\partial \RR^{i}_{j})$, etc.~
In these situations, the function $\tau \mapsto \rho(\tau)$ determining the domain of $a,...,e$
will often depend on $\tau$ through $\J$, i.e., $\rho(\tau) = \widehat{\rho}(\J(\tau))$. \parn
The structure \rref{defby} for $a,b$, etc. appears naturally in cases where these functions
can be obtained maximizing the left hand sides of Eq.s \rref{fa}, \rref{fb}, etc. by analytical means. \parn
In more complicated situations, one could consider the possibility to determine $a, b$, etc., maximising
the left hand sides of Eq.s \rref{fa}, \rref{fb}, etc. by numerical (or partially
numerical) techniques. These would give tables
of numerical maxima, to be subsequently interpolated by elementary functions
to get $a,b$, etc.~. A second possibility is
to derive the evolution equation for the maximum points of interest as function of $\tau$,
to be coupled with the other differential equations in our general framework;
this approach should work if there are no bifurcations.
\parn
Both possibilities outlined above are especially interesting for the function $a$,
since this requires the greatest accuracy; however, they will be investigated
elsewhere. \parn
iv) The fixed point $\ell_0$ in \rref{fixedp} is given by the standard iterative formula
$\ell_0 = \lim_{n \vain +\infty} l_n$, where $l_{n} :=
\alpha(0, \ep l_{n-1})$ and $l_{1}$ is chosen arbitrarily in $\Sigma$.
One can compute numerically the sequence $(l_n)$ up to a sufficiently large value
$n=N$, and then assume $\ell_0 \simeq l_N$. ({\footnote{By the standard
theory of contractions, $|\ell_0 - l_N| \leqs (\ep M)^{N-1} | l_2 - l_1 |/(1 - \ep M)$,
where $M$ is the constant in Proposition \ref{proprinc}.}}) \parn
v) Even in cases where all the other functions have known analytical expressions,
the differential equations \rref{tre} \rref{quattro} for $\em, \en$ will be
typically too difficult to be solved analytically. So, a numerical treatment will
be necessary. \parn
If we do not have analytical expressions for $\J,\R,\K$, it may be convenient
to regard Eq.s \rref{av} \rref{sistr} \rref{sistk} \rref{tre} \rref{quattro}
as a coupled system for the unknowns $\J,\R,\K,\em, \en$, to be solved numerically
on a chosen interval $[0,U)$.
\salto
\vskip 0.2cm \noindent
\textbf{3B. The "$\Np$-operation".}
Let us fix the attention on the simple situations where the functions $\overline{f}, s, ..., \HH$
have known analytical expressions and $a,b,c,d,e$ have the form \rref{defby}, depending
on known functions $\widehat{a},...,\widehat{e}$. It is not difficult to write a program of
general use for these situations, which computes the fixed point $\ell_0$ and the
functions $\J$, $\R$, $\K$, $\em$, $\en$ solving numerically the
equations \rref{fixedp} \rref{av} \rref{sistk} \rref{tre} \rref{quattro}.
From now on, the computation of $\ell_0,\J,...,\en$ by such a program,
for given $\overline{f}, ..., \widehat{e}$ (and $I_0, \te_0$, $U$), will be referred to as the
$\Np$-\textsl{operation.} Of course, the main outcomes of this operation are the solution $\J$ of the averaged system
and the function $\en$ binding $|\L(t)|$. \parn
We have written a general program for the above purpose, using the MATHEMATICA system.
Concerning Eq.\rref{quattro} for $\en$, in this program
the derivative $\partial \alpha/\partial \tau$ is expressed
via Eq. \rref{eqquattro}; the derivatives $d \J/d \tau, d \K/d \tau$ and
$d \R/d \tau$ which occur in \rref{eqquattro} and \rref{quattro} are expressed
via Eq.s \rref{av} \rref{sistr} \rref{sistk} (MATHEMATICA is also useful, in the symbolic mode,
to produce the input of the above program, i.e., the functions $\overline{f}, ..., \widehat{e}$;
this will appear from the examples of the next section).
\salto
\textbf{3C. Testing the effectiveness of the previous method: the "$\Lp$-operation".}
By the $\Lp$-operation we mean, essentially, the computation of $\L$ by direct numerical solution of the perturbed
system on $[0,U/\ep)$. To avoid misunderstandings, we stress that in
the present framework the purpose of the $\Lp$-operation is merely
to check the reliability of the estimate $| \L(t) | \leqs \en(\ep t)$
produced by the $\Np$-operation, and to prove quantitatively that
the direct solution of the perturbed system is generally much slower than $\Np$. When $U/\ep$
is very large, the $\Lp$-operation may be impossible within reasonable
times; an example will be given in the next Section (see Figure 3f,
and the explanations for it). Of course, the main
usefulness of the $\Np$-operation is just the treatment of these cases!\parn
To be precise, the $\Lp$-operation is the numerical determination
of $\J,\L, \Te$ in the following way. First, the function $\tau \in [0,U) \vain \J(\tau)$
is obtained solving the averaged system \rref{av} for $\J$; then,
the functions $t \in [0,U/\ep) \vain \L(t), \Te(t)$ are determined
solving their exact evolution equations derived from \rref{pert} \rref{av}, i.e.,
\beq \left\{\barray{ll} (d \L/ d t) (t) = f(\J(\ep t) + \ep \L(t), \Te(t)) -
\overline{f}(\J(\ep t)), & \quad\L(0) = 0~, \\
(d \Te/ d t)(t) = \om(\J(\ep t) ) +\ep g(\J(\ep t) + \ep \L(t),\Te(t)),
& \quad\Te(0) = \te_0~. \farray \right. \label{pertel} \feq
It is easy to write a MATHEMATICA program that computes numerically $\J,\L, \Te$ for given
$f, g, \omega$, $I_0, \te_0$~.
\parn
When the $\Lp$-operation can performed within reasonable times, it can be used to test the $\Np$-procedure
along these lines: \parn
i) one compares the graph of the estimator
$\en$ (an $\Np$-output) with the graph of the function $|L|$
(an $\Lp$- output); \parn
ii) one also compares the CPU times $\TT_{\Np}$,
$\TT_{\Lp}$ for the two operations. \parn
These tests are presented in the next section; they are based on the programs mentioned
here and in paragraph 3B.
In most examples, the estimator
$\en$ practically coincides with the envelope of the rapidly oscillating graph
of $| \L |$; furthermore, $\TT_{\Np}$ is generally smaller than $\TT_{\Lp}$ by one or
more orders of magnitude.
\section{Examples.}
\label{seces}
In any example we consider, the initial condition for the angle is
always
\beq \te_0 := 0~. \feq
Given $f$, $g$ and $\omega$, the functions $\overline{f}, s,..., \GG$, $\HH$ and $\rho, a,...,e$
are computed explicitly for all $I_0$ (and $U$). After this, specific choices are made for $I_0$, $U$ and $\ep$,
and the $\Np$-operation is performed; to test the accuracy of the method,
the $\Lp$-operation is also performed and some comparisons are made, as suggested at the end of the
previous section. The results are summarized in the figures which conclude the
section.
Each figure gives the graph of the estimator $\en(\tau)$ provided by $\Np$
for $\tau \in [0,U)$; it also gives
the graph of $| \L(\tau/\ep) |$ in the same interval
(except one case, where $\Lp$ has not been possible within reasonable times). \parn
Figures referring to an example are labelled by
the same number and by a letter (so, Fig.s 1a, 1b and 1c refer
to Example 1). The legend of each figure specifies
the choices of $I_0, \ep, U$, and the CPU times $\TT_{\Np}$, $\TT_{\Lp}$ (in seconds)
in the execution of the two operations
{(\footnote{of course these times, depending on the PC employed, are merely indicative.}}).
\parn
In the chosen examples, one derives simple analytical expressions for
the functions $\J,\R,\K$ but not for $\em, \en$. However, with the view of a
general comparison between the $\Np$- and $\Lp$-operations, all examples
have been treated by the general MATHEMATICA programs mentioned in paragraphs 3B-3C, which
solve numerically all the differential equations involved. Therefore, the reported
times $\TT_{\Np}$, $\TT_{\Lp}$ include contributions from the determination of $\J,\R,\K$.
In any case, the analytical expressions of these functions are written for completeness. \parn
For each example: \parn
i) the auxiliary functions $s, ..., \HH, \rho, a, ..., e$ are reported in a table. All the related
computations are analytical; the most lengthy have been performed using MATHEMATICA in the symbolic mode. \parn
ii) The function $\rho$ always gives the distance of $\J(\tau)$ from the boundary of
the actions space $\Lambda$. \parn
iii) Some details on the computation of the functions $a$ and $b, c$ are given in the
Appendices \ref{apol} and \ref{apoll}, respectively. The expressions for $d, e$ follow trivially from the
ones for $\GG, \HH$ in the corresponding tables. \parn
\salto
\textbf{Example 1: the van der Pol equation.} This is a system of the form \rref{pert}
for $(\I, \Te)$, with
\beq d := 1~, \qquad \Lambda := (0,+\infty)~, \qquad
\om(I) := -1~, \feq
$$ f(I,\te) := I (1 - {I \over 2}) - I \cos(2 \te) + {I^2 \over 2} \cos(4 \te),~~
g(I,\te) := {1 - I \over 2} \sin(2 \te) - {I \over 4} \sin(4 \te)~. $$
The functions $\x := \sqrt{2 \I} \cos \Te$, $\vel := \sqrt{2 \I} \sin \Te$ fulfil
the equations
${\dot \x} = \vel$, ${\dot \vel} = - \x - \ep~(\x^2 - 1) \vel$,
yielding the familiar van der Pol equation ${\ddot \x} + \x + \ep~(\x^2 - 1) {\dot \x} =0$.
It is found that
\beq \overline{f}(I) = I (1 - {I \over 2})~;  \feq
the auxiliary functions
$s,v,..., \HH$ of paragraph 2B are reported in Table 1
({\footnote{We note that the domain $\Gam$ of $\GG, \HH$ is made of pairs
$(I, \delta I)$ as indicated in Table 1. In all the other
examples, $\Gam$ can be read as well from the tables.}}). \parn
The averaged system \rref{av}
has the solution
\beq \J(\tau) = {2 I_0 \over I_0 + (2 - I_0) \, e^{-\tau} } \feq
for $\tau \in [0,+ \infty)$, tending to $2$ for $\tau \vain +\infty$\,: this long time behavior is the manifestation,
in the averaging approximation, of the well known limit cycle of the van der Pol equation
($\J(\tau)$ also exists for some or all $\tau < 0$, but we are not interested in this fact).
\begin{table}
\textbf{Table 1. Auxiliary functions for Example 1.} \vskip 0.3cm \hrule
\vskip 0.2cm
For $I \in (0,+\infty)$, $\te \in \bT$ and $\delta I \in (-I,+\infty)$: \vskip 0.1cm \noindent
$ s(I,\te) =  \dd{I \over 8} \Big(4 \sin(2 \te) - I \sin(4 \te) \Big)~,~\label{eqv}
\quad
v(I,\te) = - \dd{I \over 32}
\Big( 8 - I - 8 \cos(2 \te) + I \cos(4 \te) \Big)~, $ \parn
$ p(I,\te) = \dd {I \over 8} \,\Big((4 - 2 I - I^2) \sin(2 \te) + I (I - 4) \sin(4 \te)
+ I^2 \sin(6 \te)\Big), \qquad  \overline{p}(I) = 0~, $ \parn
$ q(I,\te) = - \dd{I \over 32} \, \Big(16 - 10 I + 2 I^2 -(16 - I^2) \cos(2 \te) + I (10 - 2 I) \cos(4 \,\te)
- I^2 \cos(6 \te) \Big)~, $ \parn
$ w(I,\te) = - \dd {I \over 96} \, \Big( 24 - 24 I - I^2 - 6 (4 - 2 I - I^2) \cos(2 \te)
+ 3 I (4 - I) \, \cos(4 \te)
- 2 I^2 \cos(6 \te)~ \Big)~, $ \parn
$ u(I,\te) = - \dd{I \over 128} \,\Big( 64 - 120 I + 36 I^2  + I^3 + (- 64 + 64 I + 50 I^2 - 12 I^3) \cos(2 \te) + $ \parn
$ + 4 I (14 - 17 I - I^2) \cos(4 \te) + 6 I^2 (-3 + 2 I) \cos(6 \te) + 3 I^3 \cos(8 \te) \Big)~, $ \parn
$ \MM(I) = - 1 + I - \dd{1 \over 2} I^2~, \qquad
\GG(I, \delta I) = 0, \qquad \HH(I,\delta I) = -1~. \label{eqgh} $
\vskip 0.3cm \noindent
For $\tau \in [0,U)$, $\rho(\tau) := \J(\tau)$.
\vskip 0.3cm \noindent
For $\tau \in [0,U)$ and $r \in [0,J(\tau))$~: \vskip 0.1cm \noindent
$ a(\tau,r) :=  \dd{1 \over 8} \Big(-2 + 10 (J + r)^2 +
(J + r)^4 + 2(1 + 2 (J + r)^2)^{3/2}\Big)^{1/2}_{J = \J(\tau)}~, $ \vskip 0.1cm \noindent
$ b(\tau,r) := \dd{1 \over 96} \Big(
120 J^6 + 12 J^5 (23 + 56 r) + 3 J^4 (192 + 474 r + 517 r^2) + \label{espb} $ \parn
$ + 12 J^3 r (72 + 180 r + 157 r^2)
+ 6 J^2 r^2 (372 + 530 r  +  231 r^2\!)
+ 12 J r^3 (216  + 213 r + 46 r^2\!) + $ \parn
$ + r^4 (1404   +  690 r  +  91 r^2\!) \Big)^{1/2}_{J = \J(\tau)}~, $ \vskip 0.1cm \noindent
$ \hspace{-0.0cm} c(\tau,r) := \dd{1 \over 384}
\Big(6512 J^8 + 24 J^7 (671 + 2096 r) + 24 J^6 (1693 + 5484 r + 6956 r^2) \label{espc}
+  8 J^5 \times $ \parn
$ \times (1812 + 31188 r + 39375 r^2 + 38726 r^3)
  + 12 J^4 (768 \!+ \!4436 r \!+ \!61358 r^2 \!+ \!37966 r^3 \!+ \!29997 r^4)
+ 8 J^3 r (4680 + 39948 r + 125584 r^2 + 62193 r^3 + 35046 r^4)
+ 12 J^2 r^2 (1824 + $ \parn
$ + 52152 r + 61180 r^2 + 37311 r^3 + 12021 r^4) +
J r^3 (119808 + 445536 r + 425592 r^2 + 210995 r^3 + 41976 r^4)
+ 4 r^4 (21600 + 33024 r + 30127 r^2 + 10383 r^3 + 1377 r^4)
\Big)^{1/2}_{J = \J(\tau)} , $ \vskip 0.1cm \noindent
$ d(\tau, r) := 0~, \qquad \qquad e(\tau,r) := 1~. \label{espde} $
\vskip 0.1cm \hrule
\end{table}
The Cauchy problems
\rref{sistr}, \rref{sistk} for the unknown real functions $\R, \K$ have solutions
\beq \R(\tau) = {4 e^{-\tau} \over (I_0 + (2 - I_0) \, e^{-\tau})^2 }~, \qquad \K(\tau)=0~ \feq
for $\tau \in [0,+\infty)$. From now on, $\tau$ is confined to an interval $[0,U)$
(and, of course, $U$ will be chosen finite in the subsequent
numerical computations). \parn
Our next step is to construct functions $\rho,a,...,e$
as in paragraph 2C; these are also reported in Table 1
({\footnote{The functions $b, c$ constructed in this way could be
replaced by appropriate, simpler majorants reducing the "confidence interval" $[0, \J(\tau))$ for $r$;
for example, one could redefine $\rho(\tau) := \min(\J(\tau), 1/10)$ and infer
upper bounds for $b, c$ by means
of the inequalities $r^k \leqs r/10^{k-1}$, for $k=2,3,...$, holding for $r \in [0,\rho(\tau))$.
These upper bounds are fairly simple, since they depend linearly on $r$; of course,
their use is correct if one checks a posteriori that
$0 < \ep \,\en(\tau) < \min(\J(\tau), 1/10)$ for all $\tau \in [0,U)$. However,
to perform the $\Np$-operation in all cases presented in the figures
we have used directly the complicated expressions in Table I, since these
are easily handled by MATHEMATICA.}}).
All the functions $a,..., e$ are $C^{\infty}$ in $(\tau,r)$, and non decreasing in $r$;
they have the form $a(\tau, r) = \widehat{a}(\J(\tau), r)$,
$b(\tau,r) = \widehat{b}(\J(\tau),r),...,
e(\tau,r) = \widehat{e}(\J(\tau),r)$, where $\widehat{a}, \widehat{b}, \widehat{c}$ are read from
the Table and $\widehat{d}:= 0$, $\widehat{e} := 1$ everywhere; this corresponds
to a special case of Eq. \rref{defby}. Similar remarks could be made for the other Examples, but will be
no longer repeated.
 \parn
\salto
\textsl{Comments on this example and the figures.} Figures 1a, 1b, 1c refer to the initial data
$I_0 = 1/2$ or $I_0 = 4$, one below and the other above the critical value $I = 2$
(i.e., the limit cycle in the averaging approximation); $U$
is $10$ or $200$. The ratio $\TT_{\Np}/\TT_{\Lp}$
is between 1/150 and 1/40, in the three cases. Due to the limit cycle,
one expects $|L(\tau/\ep)|$ to be bounded on the whole interval $[0,+\infty)$; this fact is
reproduced very well by our estimator $\en(\tau)$, that appears to approach a constant
value for large $\tau$ (see in particular Figure 1c).
\vskip 0.2cm \noindent
\textbf{Example 2: a case with action-dependent frequency.} We choose
\beq d=1~, \qquad \Lambda := (0, + \infty)~, \qquad \om(I) := I~, \feq
$$ f(I,\te)  := \cc I^2 (1 - \cos (2 \te))~,
\qquad g(I,\te) := \cc I^2 (1 + \cos(2 \te))~, \qquad \cc \in \{\pm 1\}~. $$
It is
\beq \overline{f}(I) = \cc I^2~, \feq
and the auxiliary functions
$s,v,..., \HH$ are reported in Table 2. Let us comment on the vanishing of $\omega$ for $I \vain 0$. Our framework
shows this "resonance" to be false: in fact, even though Eq.s \rref{thef} \rref{st} \rref{qw}
for $s,v,w$ contain a factor $1/\omega$, in this case none of these functions is singular for $I \vain 0$,
since $f,g$ vanish in this limit more rapidly than $\omega$.\parn
The averaged system \rref{av} is fulfilled with
\beq \J(\tau) = {I_0 \over 1 - \cc \tau I_0}~~\mbox{for $\tau \in [0, W_{\cc, I_0})$}, \quad
W_{\cc, I_0} := \left \{ \barray{llll} 1/I_0 & \mbox{if~ $\cc =+1$}, \\ + \infty & \mbox{if~~$\cc =-1$.}
\farray \right. \label{accord} \feq
Eq.s \rref{sistr} \rref{sistk} for
$\R, \K$ have solutions
\beq \R(\tau) = {1 \over (1 - \cc I_0 \tau)^2}~, \qquad
\K(\tau) = {\cc I^2_0 \log(1 - \cc I_0 \tau) \over 2 (1 - \cc I_0 \tau)^2} \leqs 0~\feq
on the same domain. In the sequel we assume $\tau \in [0,U)$,
with $U \leqs W_{\cc, I_0}$;
the functions $\rho$ (the same of Example 1) and $a,b,c,d,e$ are also reported in Table 2.
\begin{table}
\textbf{Table 2. Auxiliary functions for Example 2.} \vskip 0.3cm \hrule
\vskip 0.2cm
For $I \in (0,+\infty)$, $\te \in \bT$ and $\delta I \in (-I,+\infty)$ : \vskip 0.1cm \noindent
$s(I,\te) =  - \dd{\cc \over 2} I \sin (2 \te)~, \qquad v(I,\te) = - {\cc \over 4} (1 - \cos(2 \te))~,$ \parn
$p(I,\te) = - \dd{1 \over 4}\,I^2\,\Big(2 I + 4 \,I\,\cos(2 \te) + 2\,\sin(2 \te) + 2 I\,\cos(4\,\te)
- \sin(4\,\te) \Big),~~
\overline{p}(I) = - \dd{1 \over 2} I^3~,$ \parn
$q(I,\te) = - \dd{1 \over 4} \, I^2 \, \Big(2 \, \sin(2 \te) + \sin(4 \,\te) \Big)$, \parn
$w(I,\te) = - \dd{1 \over 16} I \Big( 3 - 4 \,\cos(2 \te) + 8 I \, \sin(2 \te) \, + \cos(4 \,\te)
+ 2 I \, \sin(4 \,\te) \Big)$, \parn
$u(I,\te) =  - \dd{\cc \over 32} \,I^2\,\Big( 16 \,I^2 + 10 + (40\,I^2 - 15)\,\cos(2 \te) + 40 I \sin(2 \te) + $ \parn
$+ (32 \,I^2 + 6)\,\cos(4\,\te) - 8\,I\,\sin(4\,\te) + (8\,I^2 - 1) \cos(6\,\te) - 8\,I\,\sin(6\,\te) \Big)~,$ \parn
$\MM(I) = 6 I^2~, \qquad
\GG(I, \delta I) := - \dd{1 \over 2} (3 I^2 + 3 I \delta I + \delta I^2)~,
\qquad \HH(I,\delta I) := 2 \cc$~. \parn
For $\tau \in [0,U)$, $\rho(\tau) := \J(\tau)$.
\vskip 0.3cm \noindent
For $\tau \in [0,U)$ and $r \in [0,J(\tau))$~: \vskip 0.1cm \noindent
$a(\tau, r) := \dd{1 \over 2} (\J(\tau) + r) - \K(\tau)~, \label{espaa}$ \parn
$b(\tau,r) := \dd{1 \over 8 \sqrt{2}}~\Big(50 J^4  +
    (55 + 200 r) J^3  +  (38 + 85 r + 300 r^2) J^2
+  (65 + 33 r + 200 r^2) J r + $ \parn
$ + (32 + 27 r + 50 r^2) r^2 \Big)^{1/2}_{J = \J(\tau)}~, $ \parn
$ c(\tau,r) := \dd{1 \over 16 \sqrt{2}}~\Big(4608 J^8 +
(3904+36864 r) J^7 +
(1520 + 23296 r + 129024 r^2) J^6 + (1856 \! + $
$ + 5696 r \! + 57792 r^2 \! + 258048 r^3) J^5 +
(4853 \! + 5352 r \! + 10032 r^2 \! + 76160 r^3 \! + 322560 r^4) J^4 + $ \parn
$ + (3086 \! + 7824 r \! + 11008 r^2 \! + 56000 r^3 \! + 258048 r^4)  J^3 r
+ (1862 \! + 2976 r \! + 9808 r^2 \! + $ \parn
$ + 21504 r^3 \! + 129024 r^4)  J^2 r^2
+ (1024  \! + 2312 r \! + 5440 r^2 \! + 7168 r^3 \! + 36864 r^4) J r^3  + $ \parn
$ + (512 \! + 752 r \! + 1296 r^2 \! + 1280 r^3 \! + 4608 r^4) r^4 \Big)^{1/2}_{J = \J(\tau)}~, $ \parn
$ d(\tau, r) := \dd{1 \over 2} \, (3 J^2 + 3 J r + r^2)_{J = \J(\tau)},
\qquad e(\tau,r) := 2~. $
\vskip 0.1cm \hrule
\end{table}
 \salto
\textsl{Comments on this example and the figures.} Fig.s 2a, 2b and 2c refer to the case $\cc =1$,
while Fig.s 2d and 2e refer to $\cc = -1$; the initial datum is always $I_0 = 1$.
The two cases are radically different: in fact, according to Eq. \rref{accord},
the solution $\J(\tau)$ of the averaged system diverges for $\tau \vain 1^{-}$
if $\cc = 1$, whereas for $\cc = -1$ it is defined for arbitrarily large $\tau$
and vanishes for $\tau \vain + \infty$. The figures seem to indicate a similar behaviour
for the function $\tau \mapsto |L(\tau/\ep)|$; this behaviour is reproduced very
well by our estimator $\en(\tau)$, which remains close to the envelope of $|L(\tau/\ep)|$
even for $\kappa = 1$ and $\tau$ close to $1$ (see, in particular, Fig.s 2a and 2c).
\vfill \eject \noindent
\textbf{Example 3: a truly resonant case.} Let us pass to a case where the vanishing of $\omega$ for $I \vain 0$
gives rise to singularities for $s, v, w$ and other auxiliary functions. We assume
\beq d=1~, \qquad \Lambda := (0,+\infty)~, \qquad \om(I) := I~, \feq
$$ f(I,\te)  := 1 - \cos\te~,
\qquad g(I, \te) := 0~. $$
This example is considered in \cite{Loc} \cite{Ver} to introduce the subject of
resonances; it is inspired by a two-frequency example in \cite{Arn2}.
In this case,
\beq \overline{f}(I) = 1~; \feq
the functions
$s,...., \GG,\HH$ are reported in Table 3. The averaged system \rref{av}
has the solution
\beq \J(\tau) = I_0 + \tau \label{jt} \feq
for $\tau \in [0, +\infty)$.
Eq.s \rref{sistr} \rref{sistk} for $\R$, $\K$ are very simple in this case, since
$\FF=0$ and $\overline{p}=0$;  this implies
\beq \R(\tau)=1~, \qquad \K(\tau)=0~. \feq
From now on, $\tau \in [0,U)$;
the functions $\rho, a,b,c,d,e$ are reported in Table 3.
\begin{table}[t]
\textbf{Table 3. Auxiliary functions for Example 3.} \vskip 0.3cm \hrule
\vskip 0.2cm
For $I \in (0,+\infty)$, $\te \in \bT$ and $\delta I \in (-I,+\infty)$ : \vskip 0.1cm \noindent
$ s(I,\te) =  - \dd{1\over I} \sin \te~, \qquad v(I,\te) = - {1 \over I^2}(1 - \cos \te)~,$ \parn
$ p(I,\te) = \dd{1 \over 2 I^2} (2 \sin \te - \sin(2 \te))~, \qquad ~\overline{p}(I) = 0 ~, $ \parn
$ q(I, \te) = \dd{1 \over I^3} (3 - 4 \cos \te + \cos(2 \te))~, \qquad w(I,\te) = {q(I,\te) \over 4}~, $ \parn
$ u(I, \te) = \dd{3 \over 8 I^4}~(-10 + 15 \cos \te - 6 \cos(2 \te) + \cos(3 \te))~, $ \parn
$ \MM(I) = 0~, \qquad \GG(I, \delta I) := 0~,
\qquad \HH(I,\delta I) := 0~. $
\vskip 0.3cm \noindent
For $\tau \in [0,U)$, $\rho(\tau) := \J(\tau)$.
\vskip 0.3cm \noindent
For $\tau \in [0,U)$ and $r \in [0,J(\tau))$~: \vskip 0.1cm \noindent
$a(\tau,r) :=  \dd{1 \over \J(\tau) - r}~, \qquad
b(\tau,r) := \dd{2 \over (\J(\tau) - r)^3}~, \qquad
c(\tau,r) := \dd{12 \over (\J(\tau) - r)^4}~, $ \parn
$d(\tau,r) := 0~, \qquad e(\tau,r) := 0~.$
\vskip 0.1cm \hrule
\end{table}
\salto
\textsl{Comments on this example and the figures.} The resonance for $I \vain 0^{+}$ could be expected to give
problems for initial data $I_0$ close to zero (these problems should appear mainly for small $\tau$,
since Eq. \rref{jt} for $\J$ shows a departure from the resonance as
$\tau$ grows).
As a matter of fact, the estimator $\en$ approximates well the envelope of $| \L(\tau/\ep)|$ even for small $\tau$ and
data fairly close to zero, such as $I_0 = 1/2$\,: the agreement is rather good for $\ep =10^{-2}$
(Fig.s 3a and 3b)
and very good for $\ep = 10^{-3}$ (Fig.s 3c and 3d).  \parn
The agreement between $\en$ and the envelope of $| \L |$ is very good even for $\ep = 10^{-2}$, if
we consider the larger datum $I_0 = 2$ (Fig.3e).
Fig.3f refers to the same situation on the larger interval $\tau \in [0, 200)$.
The statement on $\TT_{\Lp}$ in the legend means that
the numerical computation of $\L$ was interrupted after 240 seconds,
when the package had not yet produced a result; note that, on the contrary, the $\Np$-operation
for the same interval is very fast.
\salto
\textbf{Example 4: damped Euler's top.} We consider the system \rref{pert}, with
\beq d= 2, \qquad \Lambda := \{ I = (\ha,\ka)~|~\ha, \ka \in (0,+\infty) \},
\qquad \om(I) = \ha \ka, \label{omeq} \feq
$$ f(I,\te) := \big(- \ha (\lau + \mu \cos(2 \te)), - \ka (\lad - \mu \cos(2 \te)\big)~,
\quad g(I,\te) := \mu \sin(2 \te)~; $$
this depends on three real coefficients $\mu,\lau,\lad$ for which we assume
\beq \lau > 0~, \qquad - \lau <  \mu <  \lau~, \qquad \lad > - \lau ~. \label{assxyp} \feq
This system is related to Euler's equations for the components
$\pp, \qq, \rr$ of the angular velocity of an axially symmetric top, in presence of weak damping.
More precisely, assume that
the moment of the damping forces is a linear function of the angular velocity, and that the linear
operator expressing this dependence has a diagonal matrix $- \ep \,  \mbox{diag}(E, F, G)$
in the reference system in which the inertia operator has the form
$\mbox{diag}(A, A, C)$, with $A, C, E, F, G, \ep \in (0,+\infty)$
({\footnote{Of course, quantities like $A,...,G$, the time $t$, etc., can be treated
as real numbers, because we suppose to have fixed all the necessary physical
units.}}). Then, Euler's equations are
\beq A \dot{\pp} + (C-A) \qq \rr = - \ep E \pp~, \quad A \dot{\qq} - (C-A) \pp \rr = - \ep F \qq~, \quad
C \dot{\rr} = - \ep G \rr~. \label{eul} \feq
Given this system, we define
$\mu,\lau,\lad$ through the equations
\beq E = A (\mu + \lau)~, \qquad F = A (\lau-\mu)~, \qquad G = C(\lau + \lad)~, \feq
which imply the inequalities \rref{assxyp}. Now,
if $(\I, \Te) = (\I^1, \I^2, \Te)$ is such that $\dot \I = \ep f(\I,\Te)$ and
$\dot \Te = \om(\I) + \ep g(\I,\Te)$, the functions
\beq \pp := \I^1 \cos \Te~, \qquad \qq := \I^1 \sin \Te~, \qquad \rr := {A \over C - A} \I^1 \I^2~\feq
fulfil Euler's equations \rref{eul}. \parn
Let us return to \rref{omeq}. This implies
\beq \overline{f}(I) = (-\lau \ha, - \lad \ka)~; \feq
the functions $s, ..., \GG, \HH$ are reported in Table 4. The averaged system
has the solution
\beq \J^i(\tau) = I^{i}_0 \, e^{- \lambda_i \tau}
\qquad(i = 1, 2)~ \label{jit} \feq
for $\tau \in [0, +\infty)$. Eq.s \rref{sistr} \rref{sistk}
for the $2 \times 2$ matrix function $\R$ and for the 2-component function $\K$ have the solutions
\beq \R(\tau)= \mbox{diag}(e^{- \lambda_1 \tau}, e^{- \lambda_2 \tau})~, \qquad \K(\tau) = (0,0)~. \feq
From now on, $\tau$ is confined as usually to an interval $[0,U)$. The functions $\rho,a,...,e$ for this
example are reported in Table 4; the length of the expressions of $b, c$ is mainly due to the need
for covering all possible values of $\lambda_1, \lambda_2, \mu$.
\salto
\textsl{Comments on this example and the figures.} In this case the main difficulty
is the fact, following from \rref{jit} \rref{assxyp}, that
$\J^1(\tau)\, \J^2(\tau) = I^{1}_{0} \,I^{2}_{0} e^{- (\lambda_1 + \lambda_2) \tau}$
is small for large $\tau$. On the other hand, $\omega(I)$ vanishes for $I^1 \,  I^2 \vain 0$, and in this limit
many auxiliary functions diverge; so, the averaged system falls exponentially into a resonance. \parn
In this situation one expects a rapid growth of $|\L|$, which is in fact confirmed by
Fig.s 4a-4d; the same figures show that our estimator $\en(\tau)$ approximates
well the envelope of $|L(\tau/\ep)|$ on $[0,U]$, when $U$ is of the order of the unity.
In Fig. 4d, a good agreement between $| \L(\tau/\ep) |$ and
$| \en(\tau) |$ is attained on the longest interval among the four pictures
(namely, for $\tau \in [0,3)$). This is because
we take, simultaneously, the largest value for $I^1_0 \,I^2_0$
and the lowest values for $\ep$ and $\lambda_1 + \lambda_2$.
\begin{table}[b]
\textbf{Table 4. Auxiliary functions for Example 4.} \vskip 0.3cm \hrule
\vskip 0.2cm
For $I = (I^1, I^2) \in (0,+\infty)^2$, $\te \in \bT$ and $\delta I = (\delta I^1,
\delta I^2) \in (-I^1,+\infty) \times (-I^2, +\infty)$ : \vskip 0.1cm \noindent
$ s(I,\te) =  \dd{\mu \over 2} \sin(2 \te) \left(-{1\over \ka}~, {1 \over \ha}\right),
\qquad v(I,\te) = \dd{\mu \over 2 \ha \ka} \sin^2 \te  \left(-{1\over \ka}, {1 \over \ha}\right)~, $ \parn
$p(I,\te) = \dd{\mu \sin(2 \te) \over 2} \left(- {\lad + \mu \cos(2 \te) \over \ka},
{\lau + 3 \mu \cos(2 \te) \over \ha}\right), \qquad
\overline{p}(I) = (0,0) ~,$ \parn
$ q(I,\te) = \dd{\mu \sin^2 \te \over 2 \ha \ka} \left( - {2 \lad + 2 \mu + \lau + \mu \cos(2 \te) \over \ka},
{\lad + 2 \mu + 2 \lau + 3 \mu \cos(2 \te) \over \ha} \right), $
\end{table}
\vfill \eject \noindent
$ w(I,\te) = \dd{\mu \sin^2 \te \over 2 \ha \ka} \left( - {\lad + \mu \cos^2 \te \over \ka},
{\lau + 3 \mu \cos^2 \te \over \ha} \right)~, \qquad u(I,\te) = \dd{\mu \sin^2 \te \over 4
  \,  \ha \ka} \times $ \parn
$ \times \Big(
- \dd{4 \lad^2 + 6 \lad \mu + 2 \lad \lau +  \mu \lau +  \mu (4 \lad + 3 \mu + \lau) \cos(2 \te) +
3 \mu^2 \cos^2(2 \te) \over \ka}~, $ \parn
$ \dd{3 \lad \mu + 2 \lad \lau + 10 \mu \lau + 4 \lau^2 + 3 \mu (\lad + 5 \mu + 4 \lau) \cos(2 \te) +
15 \mu^2 \cos^2(2 \te) \over \ha}\Big)~, $ \parn
$ \FF(I) = \left( \barray{cc} - \lau & 0 \\ 0 & - \lad \farray \right),~~\MM(I) =
\left( \barray{cc} - \lau^2 & 0 \\ 0 & - \lad^2 \farray \right),~~
\GG(I, \delta I) = 0,~~\HH(I,\delta I) = 0~. $
\vskip 0.3cm \noindent
For $\tau \in [0,U)$, $\rho(\tau) := \min(\J^1(\tau), \J^2(\tau))$.
\vskip 0.3cm \noindent
For $\tau \in [0,U)$ and $r \in [0,J(\tau))$~: \vskip 0.1cm \noindent
$a(\tau,r) := \dd{| \mu | \over 2} \left( {1 \over (\J^1(\tau) - r)^2} + {1 \over (\J^2(\tau) - r)^2} \right)^{1/2}~;$
\vskip 0.1cm \noindent
$ b(\tau,r) := | \mu |
~\dd{\big( b_{1 1} \J^1(\tau)^2 + b_{2 2} \J^2(\tau)^2 + b_1 \J^1(\tau) r + b_2 \J^2(\tau) r
+ b_0 r^2 \big)^{1/2} \over 8 (\J^1(\tau) - r)^2 (\J^2(\tau) - r)^2}
~, $
\vskip 0.1cm \noindent
$ b_{11} := 16 ( \lau^2 + \lad^2)   + \lau (12 \lad + 20 | \lad |) + 2  (\lau+\lad)  \mu +
    4  (\lau + |\lad|) | \mu | + \mu^2~, $ \parn
$ b_{2 2} := 16 ( \lau^2 + \lad^2)
+ \lau (12 \lad + 20 | \lad |) + 6  (\lau + \lad) \mu  +
    12  (\lau + | \lad |) | \mu | + 9  \mu^2~, $ \parn
$ b_{1} :=
32  (\lau^2 + \lad^2)     +
    64      \lau | \lad | + 12    (\lau + | \lad |) | \mu |  + 2    \mu^2~, $ \parn
$ b_{2} :=
32 ( \lau^2 + \lad^2)   +
    64      \lau  | \lad | + 36 (\lau + | \lad |) | \mu | + 18    \mu^2~, $ \parn
$ b_0 :=
16  (\lau^2 + \lad^2)   +
    \lau (12 \lad + 20 | \lad |) + 4  (\lau + \lad)  \mu  +
    14  (\lau + | \lad |) | \mu |  + 9   \mu^2~; $
\vskip 0.1cm \noindent
$ c(\tau,r) := | \mu |\dd{\big( c_{1 1} \J^1(\tau)^2 + c_{2 2} \J^2(\tau)^2 + c_1 \J^1(\tau) r +
c_2 \J^2(\tau) r + c_0 r^2 \big)^{1/2} \over 32 (\J^1(\tau) - r)^2 (\J^2(\tau) - r)^2}~, $
\vskip 0.1cm \noindent
$ c_{1 1} :=
1024 (\lau^4+\lad^4) + 6144 \lau^2 \lad^2 + 512 (\lau^2+ \lad^2) \lau (3 \lad + 5 | \lad |)
+ 640 (\lau^3+ \lad^3)  \mu + $ \parn
$  + 896 (\lau^3+ |\lad|^3) | \mu |
 + 1920 (\lau+\lad) \lau \lad \mu + 2688 (\lau+ |\lad|) \lau |\lad| | \mu | + 704 (\lau^2+ \lad^2) \mu^2  + $ \parn
$ + 32 \lau (17 \lad + 27 | \lad |) \mu^2
- 24 (\lau+\lad)  \mu^3 + 264 (\lau+ |\lad|) | \mu |^3 + 27 \mu^4~,$ \parn
$
c_{2 2} :=
1024 (\lau^4+ \lad^4) + 6144 \lau^2 \lad^2 + 512 (\lau^2+ \lad^2) \lau (3 \lad + 5 | \lad |)
+ 384 (\lau^3+\lad^3) \mu  + $ \parn
$ + 1408 (\lau^3+ | \lad |^3) | \mu |
+ 1152 (\lau+\lad) \lau \lad \mu + 4224 (\lau + |\lad|) \lau |\lad| | \mu |
+ 2816 (\lau^2+\lad^2) \mu^2 + $ \parn
$  + 32 \lau  (21 \lad + 155 | \lad |) \mu^2
+ 120 (\lau+ \lad) \mu^3 + 1800 (\lau+|\lad|) |\mu|^3 + 675 \mu^4~, $ \parn
$c_1 :=
2048 (\lau^4+ \lad^4) + 12288 \lau^2 \lad^2 + 8192 (\lau^2+ \lad^2) \lau  | \lad | +
  3072 (\lau^3+ |\lad|^3) | \mu | + $ \parn
$ + 9216 (\lau+ |\lad|) \lau |\lad| | \mu | + 1408 (\lau^2+ \lad^2) \mu^2 +  2816 \lau | \lad | \mu^2
+ 576 (\lau+|\lad|) | \mu |^3 + 54 \mu^4 ~, $ \parn
$
c_2 :=
2048 (\lau^4+ \lad^4) + 12288 \lau^2 \lad^2 + 8192 (\lau^2+ \lad^2) \lau  | \lad |
+ 3584 (\lau^3+ |\lad|^3) | \mu | + $ \parn
$  + 10752(\lau+ |\lad|) \lau |\lad| | \mu | + 5632 (\lau^2+ \lad^2) \mu^2  +  11264 \lau | \lad | \mu^2
+ 3840 (\lau+|\lad|) | \mu |^3 + 1350 \mu^4 ~, $ \parn
$
c_0 :=
1024 (\lau^4+\lad^4)  + 6144 \lau^2 \lad^2 + 512 (\lau^2+ \lad^2) \lau (3 \lad + 5 | \lad |)
+ 512 (\lau^3+ \lad^3) \mu  + $ \parn
$ + 2048 (\lau^3 + |\lad|^3) | \mu | + 1536 (\lau+\lad) \lau \lad \mu
  + 6144 (\lau+|\lad|)\lau |\lad| | \mu | + 2816 (\lau^2+ \lad^2) \mu^2 + $ \parn
$   + 32 \lau (19 \lad + 157 | \lad |)  \mu^2
+ 48 (\lau+\lad) \mu^3 + 1872  (\lau+|\lad|) | \mu |^3 + 675 \mu^4~; $
\vskip 0.1cm \noindent
$ d(\tau, r) := 0~; \qquad \qquad e(\tau,r) := 0~. $ \parn
\vskip 0.1cm \hrule
\vskip 0.4cm \noindent
\begin{figure}
\parbox{3in}{
\includegraphics[
height=2.0in,
width=2.8in
]%
{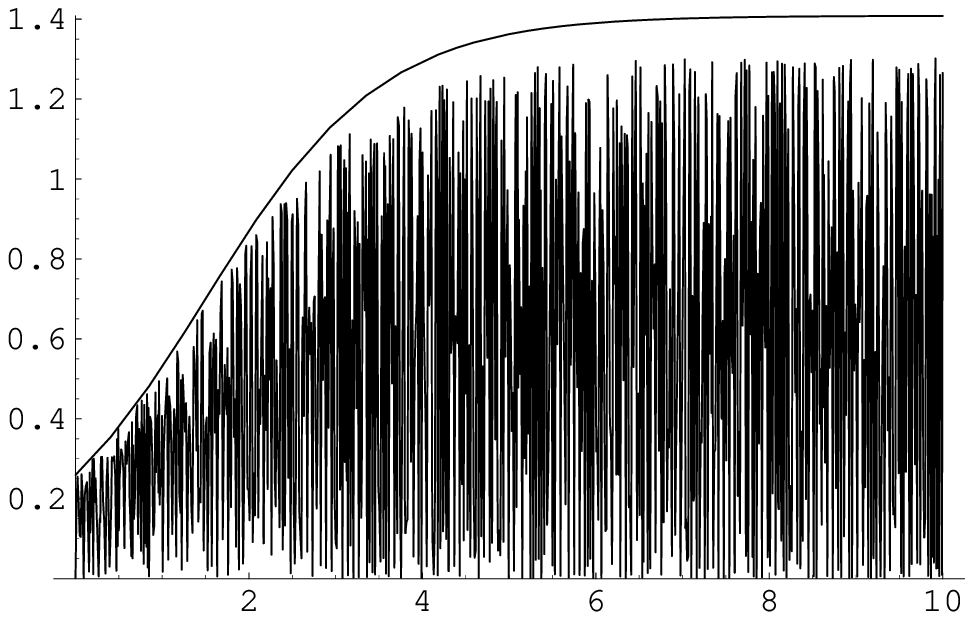}%
\parn
{\textbf{Figure 1a.~} $I_0=1/2$, $\ep=10^{-2}$, $U = 10$. \parn
Graphs of $\en(\tau)$ and $|L(\tau/\ep)|$. $\TT_{\Np} = 0.062 s$,\parn
$\TT_{\Lp} = 3.2 s$. \parn}
\label{f1a}
}
\hskip 0.4cm
\parbox{3in}{
\includegraphics[
height=2.0in,
width=2.8in
]%
{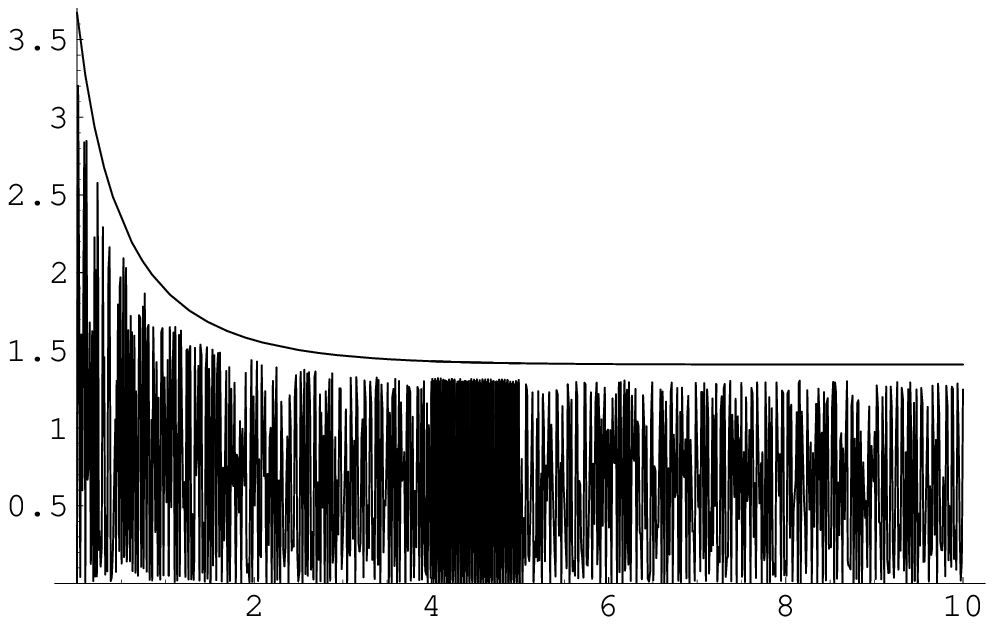}%
\parn
{\textbf{Figure 1b.~} $I_0=4$, $\ep=10^{-2}$, $U=10$. \parn
Graphs of $\en(\tau)$ and $|L(\tau/\ep)|$. $\TT_{\Np} = 0.078 s$,
$\TT_{\Lp} = 3.0 s$. \parn}
\label{f1b}
}
\parbox{3in}{
\includegraphics[
height=2.0in,
width=2.8in
]%
{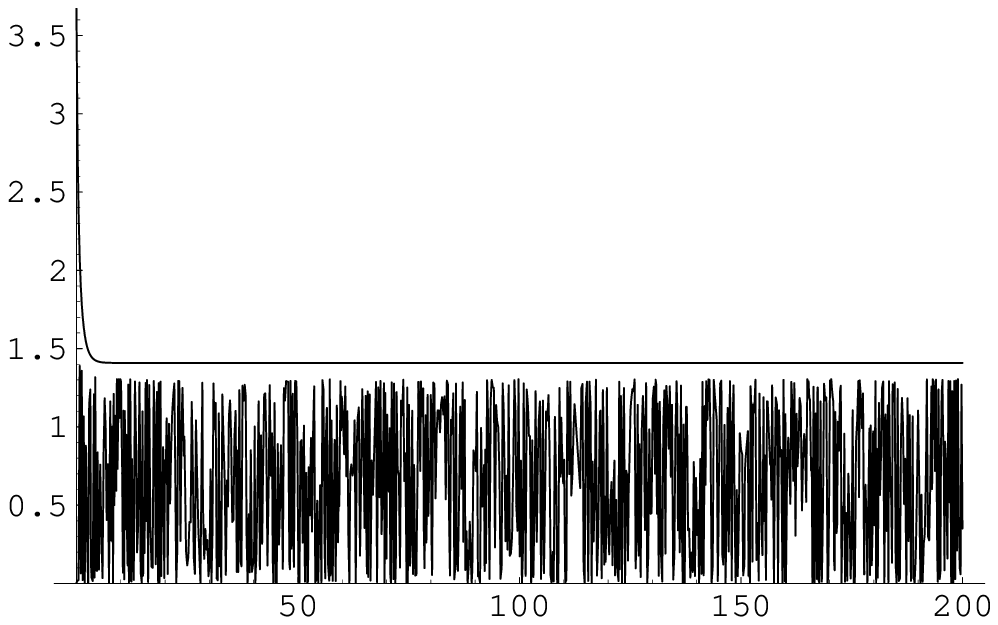}%
\parn
{\textbf{Figure 1c.~} $I_0=4$, $\ep=10^{-2}$, $U = 200$. \parn
Graphs of $\en(\tau)$ and $|L(\tau/\ep)|$. $\TT_{\Np} = 0.45 s$,\parn
$\TT_{\Lp} = 67 s$. \parn}
\label{f1c}
}
\hskip 0.4cm
\parbox{3in}{
\includegraphics[
height=2.0in,
width=2.8in
]%
{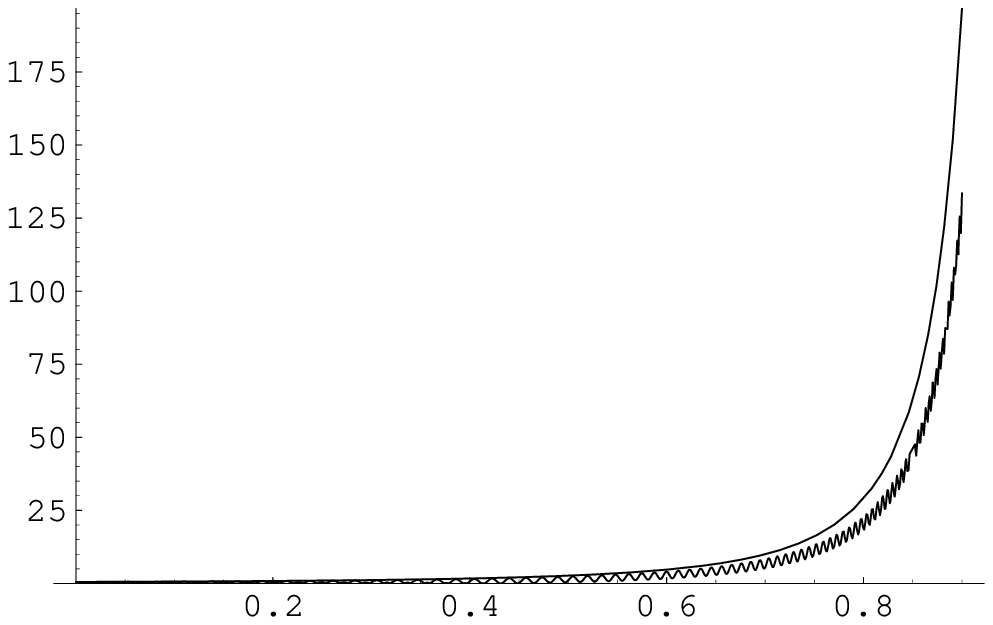}%
\parn
{\textbf{Figure 2a.~} \hskip -0.2cm $\cc \!= \! 1, \! I_0 \! =\! 1,  \! \ep\! =\! 10^{-2}, \! U \! = \! 0.9.$
Graphs of $\en(\tau)$, $|L(\tau/\ep)|$ (note that $\J(\tau) \! \vain \! + \infty$
for $\tau \! \vain \! 1^{-}$). $\TT_{\Np} \! = \! 0.032 s,
\TT_{\Lp} \! = \! 0.36 s$. \parn}
\label{f2a}
}
\parbox{3in}{
\includegraphics[
height=2.0in,
width=2.8in
]%
{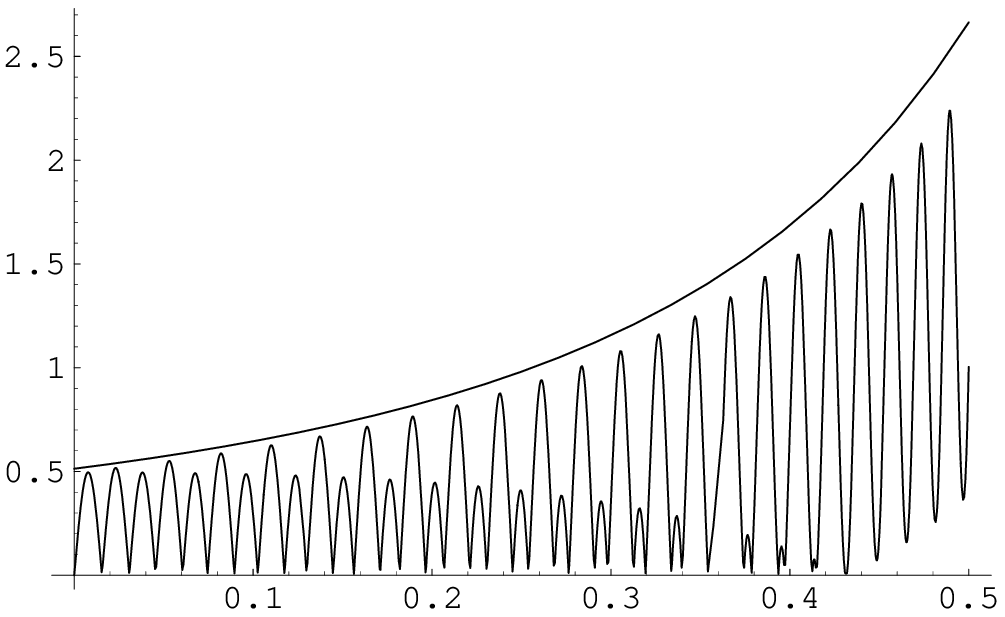}%
\parn
{\textbf{Figure 2b.~} \hskip -0.2cm $\cc \!= \! 1, \! I_0 \! =\! 1,  \! \ep\! =\! 10^{-2}$
(as in Fig.2a).
Graphs of $\en(\tau)$, $|L(\tau/\ep)|$ in a detailed view, for $\tau \in [0, 0.5]$. \parn}
\label{f2b}
}
\hskip 0.4cm
\parbox{3in}{
\includegraphics[
height=2.0in,
width=2.8in
]%
{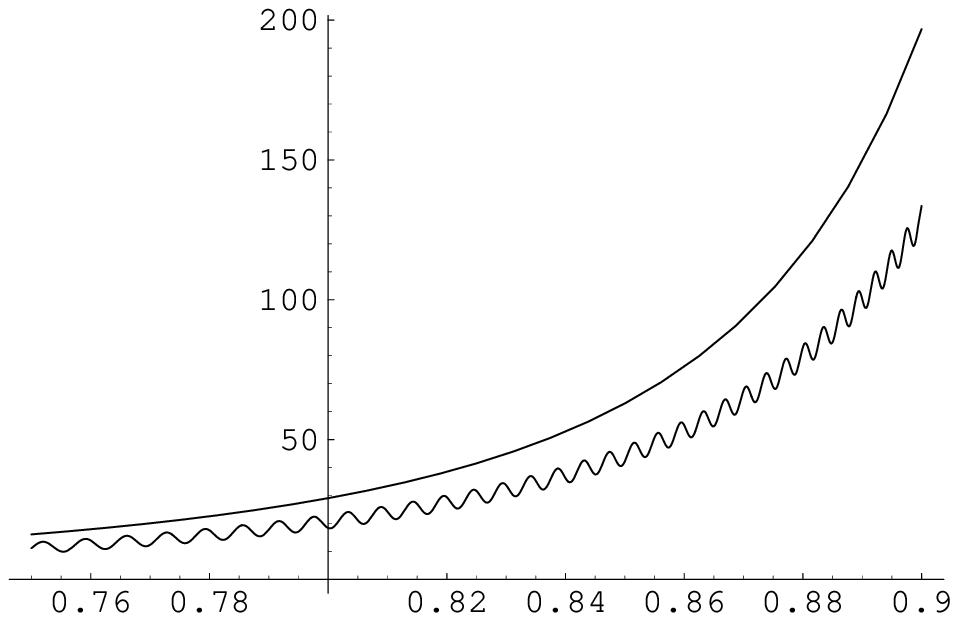}%
\parn
{\textbf{Figure 2c.~} \hskip -0.2cm $\cc \!= \! 1, \! I_0 \! =\! 1,  \! \ep\! =\! 10^{-2}$
(as in Fig.2a).
Graphs of $\en(\tau)$, $|L(\tau/\ep)|$ in a detailed view, for $\tau \in [0.75, 0.9]$. \parn}
\label{f2c}
}
\end{figure}
\begin{figure}
\parbox{3in}{
\includegraphics[
height=2.0in,
width=2.8in
]%
{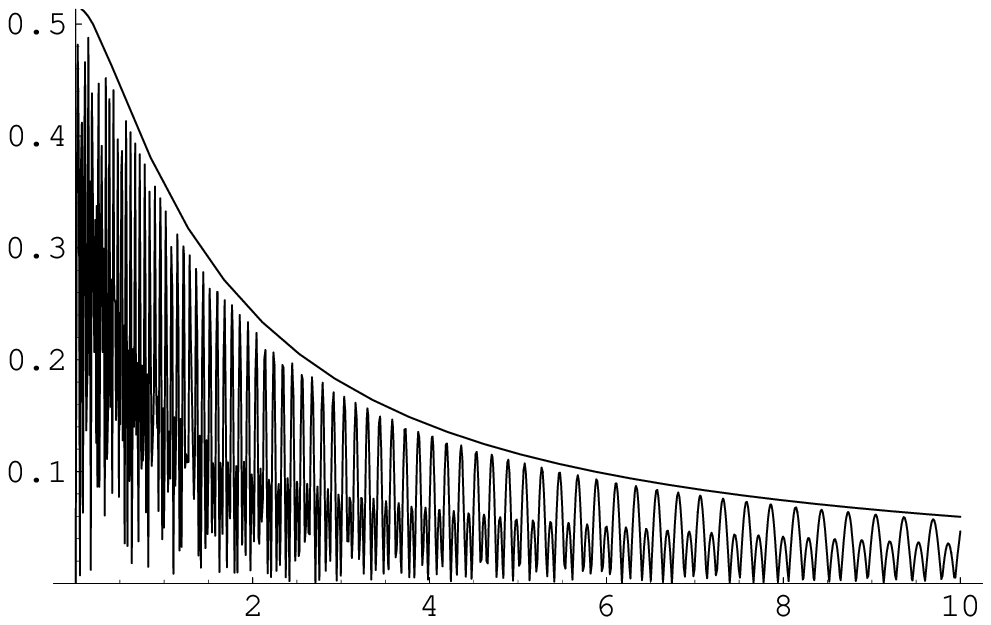}%
\parn
{\textbf{Figure 2d.~} \hskip -0.2cm $\cc \!= \! -1, \! I_0 \! =\! 1,  \! \ep\! =\! 10^{-2}, \! U \! = \! 200.$
$\TT_{\Np} \! = \! 0.078 s,
\TT_{\Lp} \! = \! 0.58 s$. Graphs of $\en(\tau)$, $|L(\tau/\ep)|$ in a detailed view,
for $\tau \in [0,10]$. \parn \parn}
\label{f2d}
}
\hskip 0.4cm
\parbox{3in}{
\includegraphics[
height=2.0in,
width=2.8in
]%
{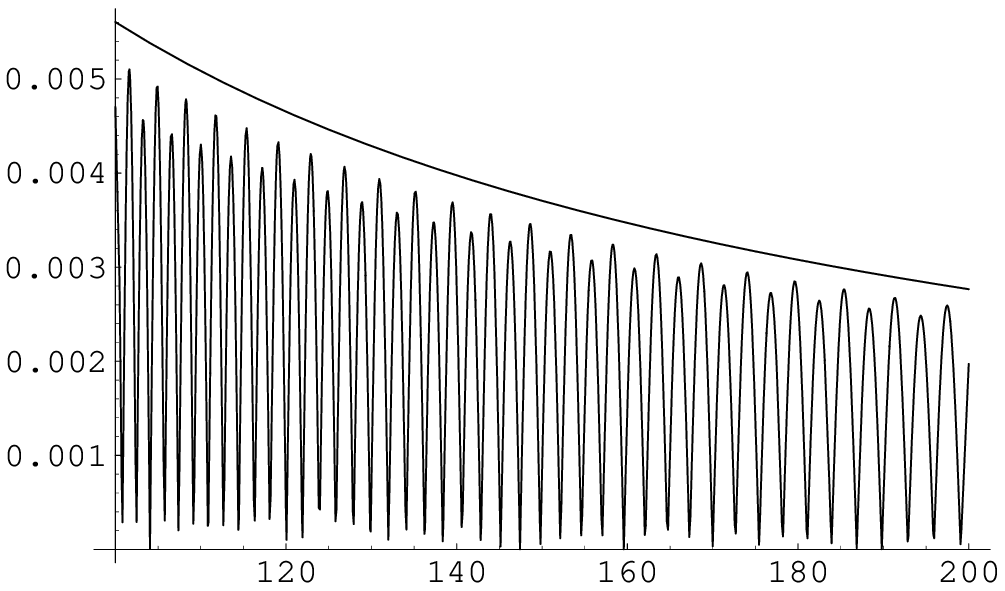}%
\parn
{\textbf{Figure 2e.~} \hskip -0.2cm $\cc \!= \! -1, \! I_0 \! =\! 1,  \! \ep\! =\! 10^{-2}$
(as in Fig.2d). Graphs of $\en(\tau)$, $|L(\tau/\ep)|$ in a detailed view,
for $\tau \in [100,200]$. \parn}
\label{f2e}
}
\parbox{3in}{
\includegraphics[
height=2.0in,
width=2.8in
]%
{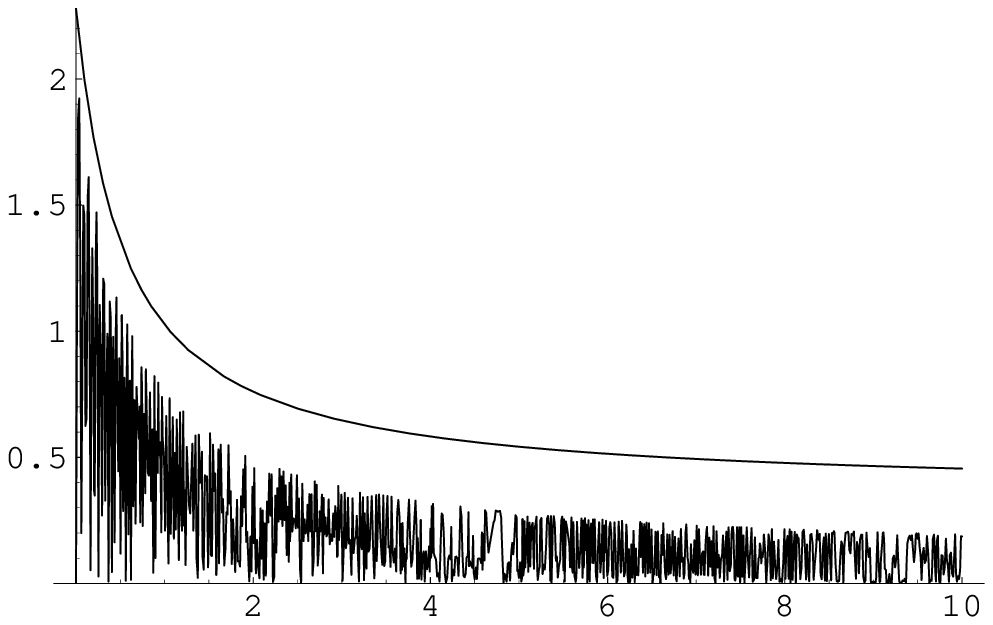}%
\parn
{\textbf{Figure 3a.~} $I_0=1/2$, $\ep=10^{-2}$, $U=10$. \parn
Graphs of $\en(\tau)$ and $|L(\tau/\ep)|$. $\TT_{\Np} = 0.23 s$, $\TT_{\Lp} = 0.95 s$. \parn \parn}
\label{f3a}
}
\hskip 0.6cm
\parbox{3in}{
\includegraphics[
height=2.0in,
width=2.8in
]%
{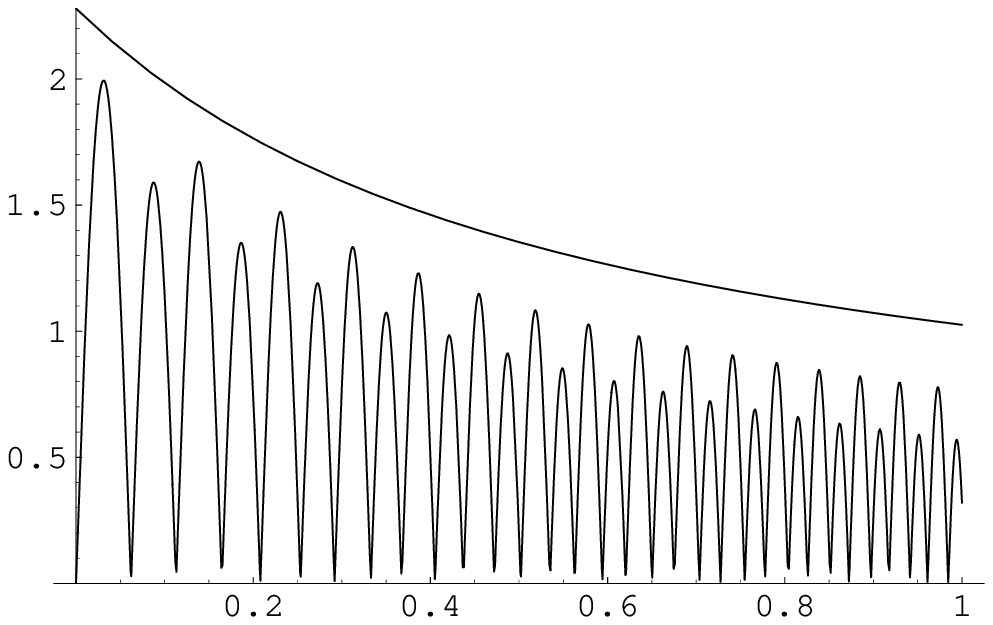}%
\parn
{\textbf{Figure 3b.~} $I_0=1/2$, $\ep=10^{-2}$ (as in Fig.3a).
Graphs of $\en(\tau)$ and $|L(\tau/\ep)|$ in a detailed view, for $\tau \in [0,1)$. \parn \parn \parn}
\label{f3b}
}
\hskip 0.4cm
\parbox{3in}{
\includegraphics[
height=2.0in,
width=2.8in
]%
{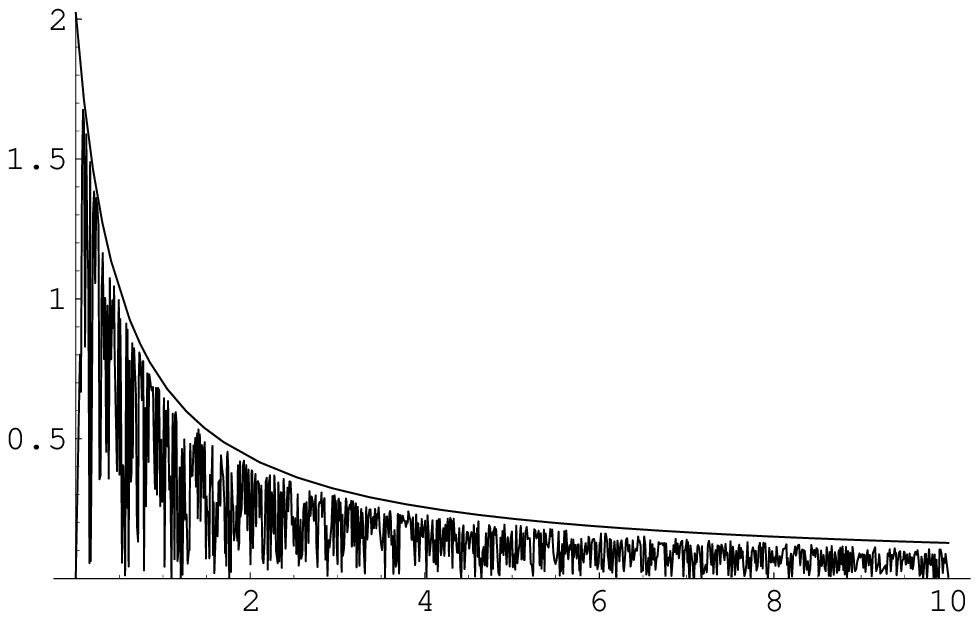}%
\parn
{\textbf{Figure 3c.~} $I_0=1/2$, $\ep=10^{-3}$, $U=10$.
Graphs of $\en(\tau)$ and $|L(\tau/\ep)|$. $\TT_{\Np} = 0.23 s$, $\TT_{\Lp} = 12 s$. \parn \parn}
\label{f3c}
}
\hskip 0.6cm
\parbox{3in}{
\includegraphics[
height=2.0in,
width=2.8in
]%
{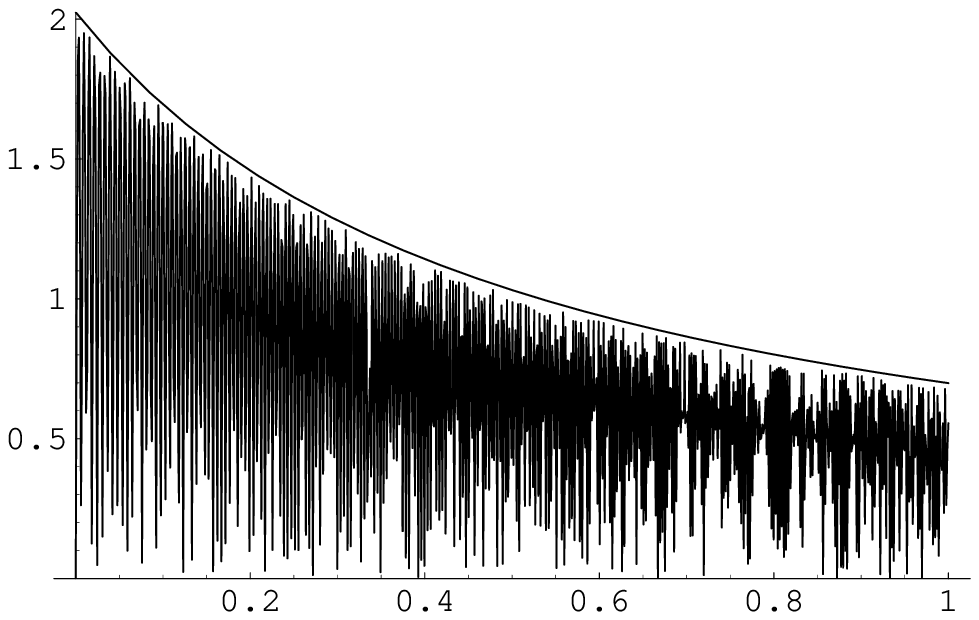}%
\parn
{\textbf{Figure 3d.} $I_0=1/2$,  $\ep=10^{-3}$ (as in Fig.3c).
Graphs of $\en(\tau)$ and $|L(\tau/\ep)|$ in a detailed view, for $\tau \in [0,1)$. \parn}
\label{f3d}
}
\end{figure}
\begin{figure}
\parbox{3in}{
\includegraphics[
height=2.0in,
width=2.8in
]%
{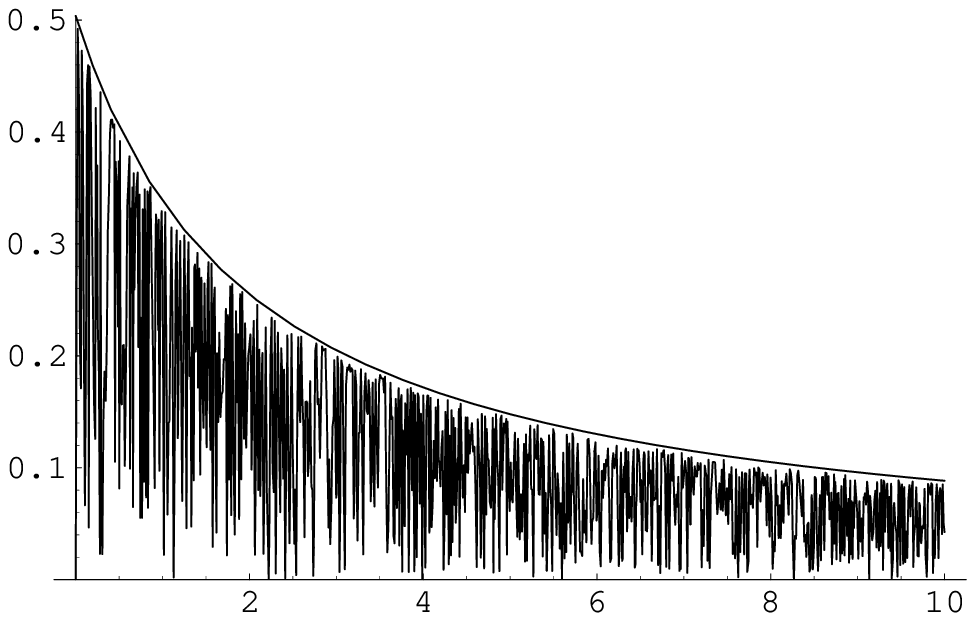}%
\parn
{\textbf{Figure 3e.~} $I_0 = 2$, $\ep = 10^{-2}$, $U=10$. \parn
Graphs of $\en(\tau)$ and $|\L(\tau/\ep)|$.
\!\!$\TT_{\Np} = 0.16 s$, $\TT_{\Lp} = 1.2 s$.\parn \parn}
\label{f3e}
}
\hskip 0.4cm
\parbox{3in}{
\includegraphics[
height=2.0in,
width=2.8in
]%
{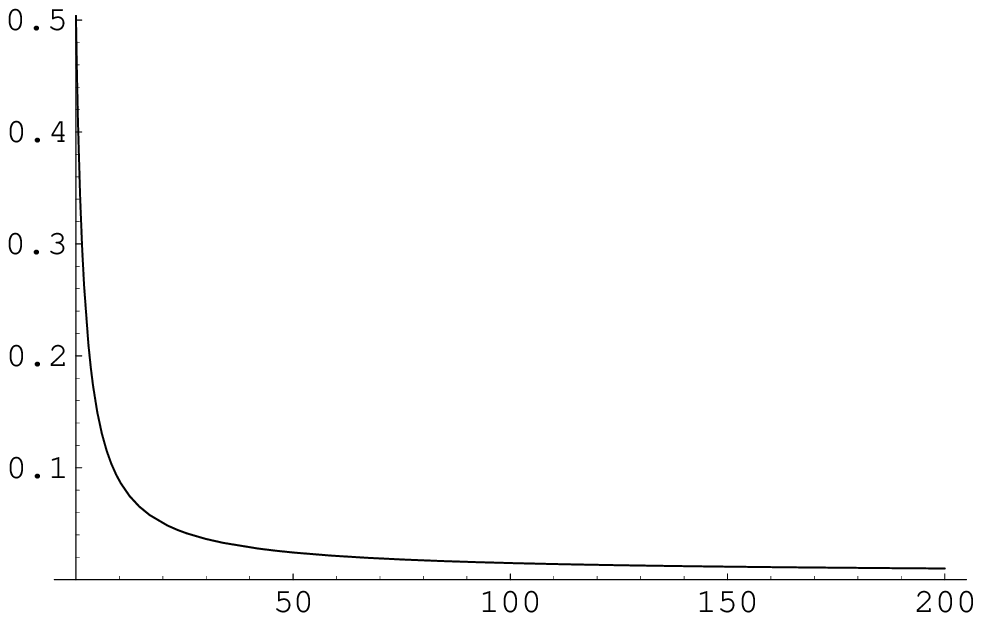}%
\parn
{\textbf{Figure 3f.~} $I_0 = 2$, $\ep = 10^{-2}$, $U=200$. \parn
Graph of $\en(\tau)$. $\TT_{\Np} = 0.28 s$, $\TT_{\Lp} > 240 s$. \parn {\,} \parn  \parn}
\label{f3f}
}
\parbox{3in}{
\includegraphics[
height=2.0in,
width=2.8in
]%
{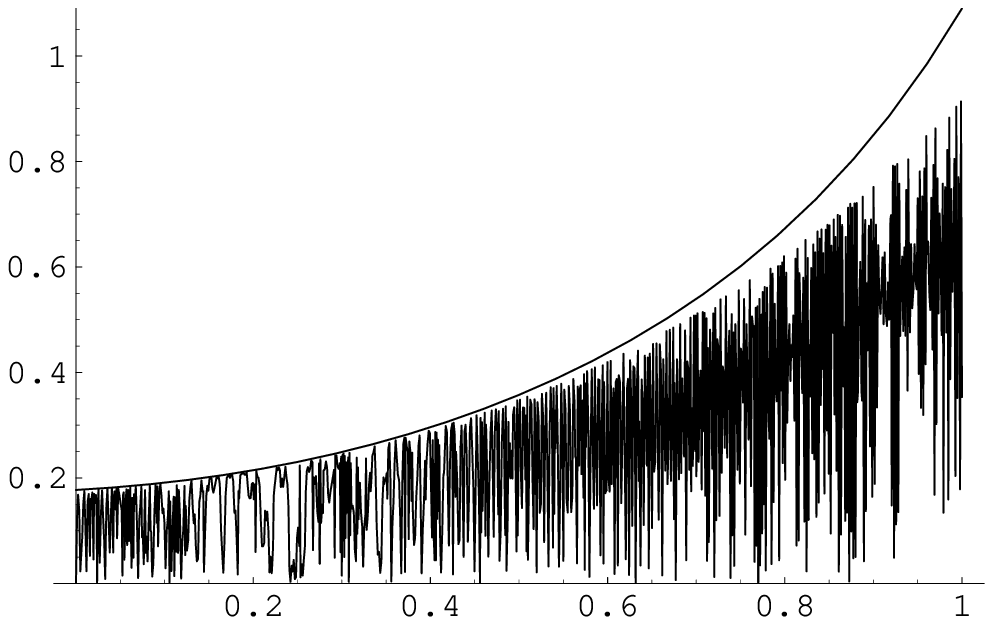}%
\parn
{\textbf{Figure 4a.~} \hskip -0.2cm $\mu \!= \! 1, \lau \! = \! 2, \lad \! = \! -1,
I^1_0 \! =\! 4$, $I^2_0 \! = \! 4,  \ep\! =\! 10^{-2}, U \! = \! 1.$
Graphs of $\en(\tau)$ and $|L(\tau/\ep)|$.
$\TT_{\Np} \! = \! 0.047 s,
\TT_{\Lp} \! = \! 1.7 s$.
\parn}
\label{f4a}
}
\hskip 0.4cm
\parbox{3in}{
\includegraphics[
height=2.0in,
width=2.8in
]%
{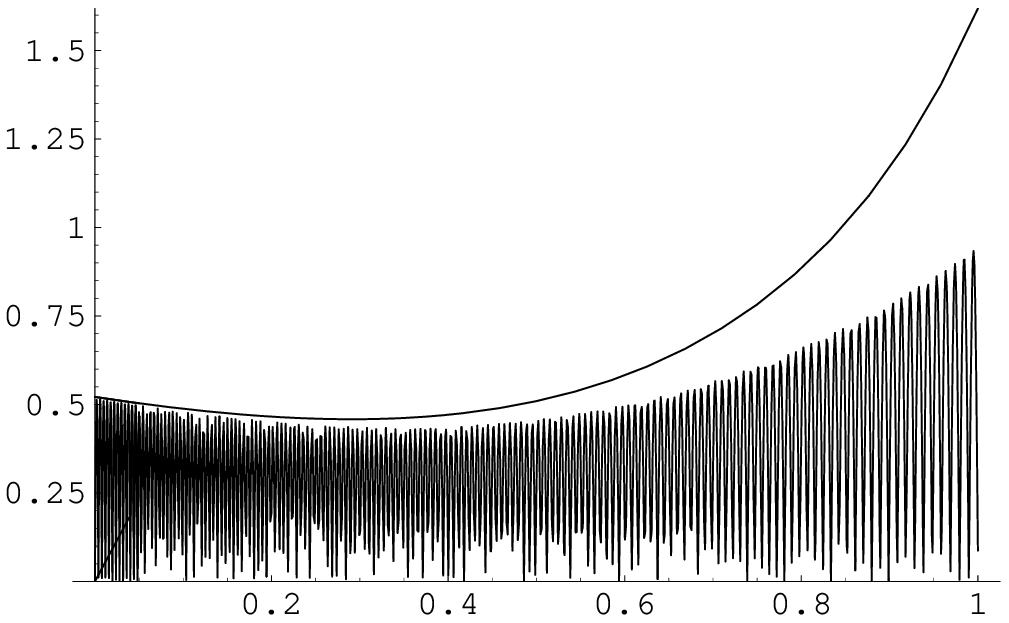}%
\parn
{\textbf{Figure 4b.~} \hskip -0.2cm $\mu \!= \! 1, \lau \! = \! 2, \lad \! = \! -1,
I^1_0 \! =\! 4$, $I^2_0 \! = \! 1,  \ep\! =\! 10^{-2}, U \! = \! 1.$
Graphs of $\en(\tau)$ and $|L(\tau/\ep)|$.
$\TT_{\Np} \! = \! 0.047 s,
\TT_{\Lp} \! = \! 0.44 s$. \parn}
\label{f4b}
}

\parbox{3in}{
\includegraphics[
height=2.0in,
width=2.8in
]%
{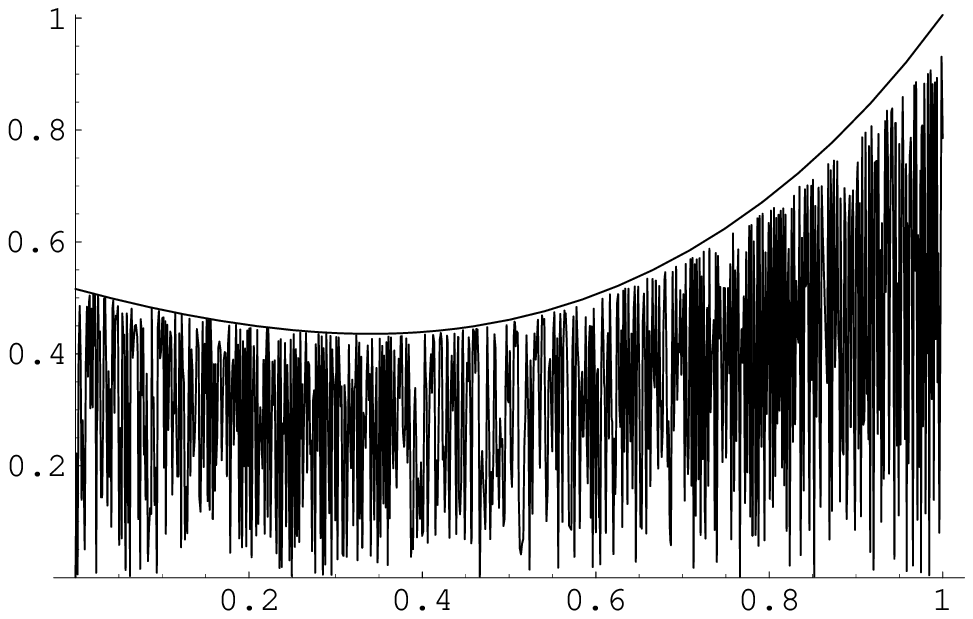}%
\parn
{\textbf{Figure 4c.~} \hskip -0.2cm $\mu \!= \! 1, \lau \! = \! 2, \lad \! = \! -1,
I^1_0 \! =\! 4$, $I^2_0 \! = \! 1, \ep\! =\! 10^{-3}, U \! = \! 1.$
Graphs of $\en(\tau)$ and $|L(\tau/\ep)|$.
$\TT_{\Np} \! = \! 0.047 s,
\TT_{\Lp} \! = \! 4.2 s$. \parn}
\label{f4c}
}
\hskip 0.6cm
\parbox{3in}{
\includegraphics[
height=2.0in,
width=2.8in
]%
{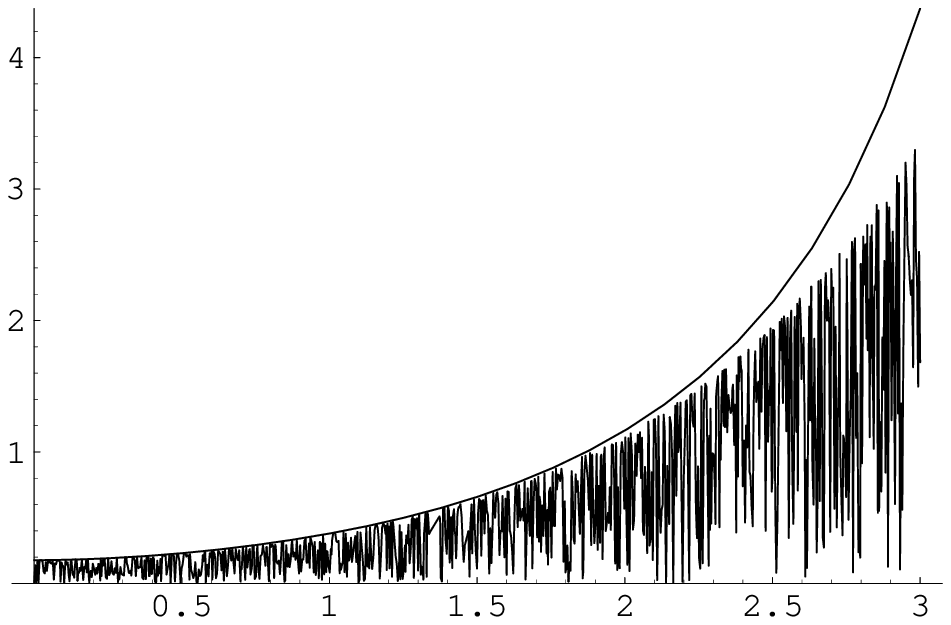}%
\parn
{\textbf{Figure 4d.~} \hskip -0.2cm $\mu \!= \! 1, \lau \! = \! 1.1, \lad \! = \! -1,
I^1_0 \! =\! 4$, $I^2_0 \! = \! 4, \ep\! =\! 10^{-3},  U \! = \! 3.$
Graphs of $\en(\tau)$ and $|L(\tau/\ep)|$.
$\TT_{\Np} \! = \! 0.046 s,
\TT_{\Lp} \! = \! 99 s$. \parn}
\label{f4d}
}
\end{figure}
\vfill \eject \noindent
\appendix
\section{Appendix. Proof of Lemma \ref{lemma1}.}
\label{aplem1}
First of all, the Cauchy problem \rref{sistr} has a (unique) solution
on $[0,U)$, and this is $C^\m$, because we have a linear differential equation for $\R$, with
a $C^{\m-1}$ matrix function $\tau \mapsto \FF(\J(\tau))$. The invertibility of $\R(\tau)$ follows
from the Wronskian identity $\det \R(\tau) = \det \R(0) \,\exp \int_{0}^{\tau} d \tau' \,
\mbox{tr} \,\FF(\J(\tau'))$
and from the initial condition $\R(0) = \uno$; the $d=1$ expression of $\R$ is obvious.
The statements on $\K$ that follow Eq. \rref{sistk} are also elementary
(as for the $C^m$ regularity, note that $\overline{p}(\J)$ is a
$C^{\m-1}$ function of $\tau$). \parn
To go on, we introduce the short-hand notations
\beq \J, \R, \K,  {d \J \over d \tau}, \mbox{etc.}
\equiv \mbox{the functions}~t \mapsto \J(\ep t)~,~\R(\ep t)~,~\K(\ep t)~,
~{d \J \over d \tau}(\ep t)~, \mbox{etc.}~; \feq
in the same spirit, for $h : \Lambda \times \bT \vain \reali^d$ and
$k : \Lambda \vain \reali^d$ we also intend
\beq h, k , k(\J)\equiv \mbox{the functions}~ t \mapsto h(\I(t),\Te(t))~,
~ t \mapsto k(\I(t))~, ~ t \mapsto k(\J(\ep t))~. \feq
In these notations, one has $\L = (\I - \J)/\ep$ and Eq.s \rref{pert}  \rref{av} imply
\beq {d \L \over d t} = {1 \over \ep} \left({d \I \over d t}
- \ep {d \J \over d \tau} \right) = f - \overline{f}(\J)~;
\label{elle} \feq
we continue dividing the argument in steps. \parn
\textsl{Step 1. One has}
\beq {d \L \over d t} = \om {\partial s \over \partial \te}
+ \ep \FF(\J) \, \L
+ {1 \over 2} \ep^2 \HH(\J,\ep \L) \, \L^2~. \label{onhas} \feq
In fact, Eq. \rref{elle} and the first Eq. \rref{thef} imply
\beq {d \L \over d t}
= \om {\partial s \over \partial \te} + \overline{f}- \overline{f}(\J)~; \feq
now, it suffices to recall that $\I = \J + \ep \L$ and use Eq. \rref{equel} with $(I, \delta I)$
replaced by $(\J, \ep \L)$~.  \parn
\textsl{Step 2. For each function $h \in C^1(\Gi \times \bT,\reali^d)$, it is}
\beq \om {\partial h \over \partial \te} =
{d h \over d t} - \ep
\Big({\partial h \over \partial I} f
+ {\partial h \over \partial \te} g \Big)~. \label{stp2} \feq
This follows easily from
\beq {d h \over d t} =
{\partial h \over \partial I} {d \I \over d t}
+ {\partial h \over \partial \te} {d \Te \over d t} = \ep {\partial h \over \partial I} f
+ {\partial h \over \partial \te} (\om +
\ep g)~. \feq
\textsl{Step 3. One has}
\beq \om {\partial s\over \partial \te} = {d s \over d t} - \ep {d w \over d t} - \ep \overline{\p}(\J) +
\ep^2(\u - \GG(\J,\ep \L) \L)~.
\label{ohs} \feq
To prove this, we note that Eq. \rref{stp2} with $h=s$ and the definition \rref{dq} of $\p$ give
\beq \om {\partial s\over \partial \te}
= {d s \over d t}  - \ep \p~.
\label{ohss} \feq
On the other hand, Eq.s \rref{qw} \rref{stp2} with
$h = \w$ and the definition \rref{em} of $\u$ imply
\beq \p = \overline{\p} +
\om {\partial \w \over \partial \te} = \overline{\p} + {d \w \over d t} -
\ep \u~; \label{dainseq} \feq
furthermore, Eq. \rref{equel0} with $(I, \delta I)$ replaced by $(\J, \ep \L)$ gives
\beq \overline{p} = \overline{p}(\J) + \ep \, \GG(\J, \ep  \L) \, \L~. \label{equel00} \feq
Inserting Eq. \rref{equel00} into \rref{dainseq}, and the result into \rref{ohss}, we get the
equality \rref{ohs}. \parn
\textsl{Step 4. One has}
\beq {d \L \over d t} - \ep \FF(\J) \L
= {d s \over d t} - \ep {d w \over d t}
- \ep \overline{\p}(\J) + \ep^2 ( \u - \GG(\J, \ep  \L) \L
+ {1 \over 2} \HH(\J,\ep \L) \L^2 )~.
\label{spt4} \feq
This follows immediately from Eq. \rref{onhas} and from the equality \rref{ohs}.
\vskip 0.1cm \noindent
\textsl{Introducing the next steps.} Eq. \rref{spt4} is an equality involving total derivatives,
and nonderivative terms proportional to $\ep$ or $\ep^2$. Our aim is to obtain an equality for $\L$
involving only total derivatives and nonderivative terms proportional to $\ep^2$;
due to the structure of the terms in $\ep$ of Eq. \rref{spt4}, this result can be achieved
using the functions $\R$ and $\K$.
In the sequel we will derive some identities involving $\R$, where
the operator $\R (d/d t) \R^{-1}$ plays a major role; inserting these relations into Eq.
\rref{spt4} (and factoring out $\R$) we will finally obtain an identity with the desired structure,
where the nonderivative terms are confined to the order $\ep^2$. \vskip 0.1cm \noindent
\textsl{Step 5. One has}
\beq {d \R^{-1} \over d t} = - \ep \R^{-1} \FF(\J)~.
\label{dermat} \feq
\textsl{For each $C^1$ function $\X : [0,U/\ep) \vain \reali^d$, $t \mapsto \X(t)$, this implies}
\beq {d \X \over d t} - \ep \FF(\J) \X = \R {d \over d t}(\R^{-1} \X)~. \label{derx} \feq
Eq. \rref{dermat} follows from the relation
\beq 0 =  {d \over d t}(\R \R^{-1}) = {d \R \over d t} \R^{-1} + \R {d \R^{-1} \over d t} =
\ep \FF(\J) + \R {d \R^{-1} \over d t}~, \feq
where, in the last passage, we have used Eq. \rref{sistr} to express $d \R/ \d t = \ep d \R/ d \tau$. \parn
Having extablished \rref{dermat}, we consider any function $\X$ as above and note that
\beq {d \X \over d t} - \ep \FF(\J) \X =
{d \X \over d t} + \R {d \R^{-1} \over d t} \X =
\R (\R^{-1} {d \X \over d t} + {d \R^{-1} \over d t} \X)~, \feq
whence Eq. \rref{derx}. \parn
\textsl{Step 6. One has}
\beq {d \L \over d t} - \ep \FF(\J) \L = \R {d \over d t}(\R^{-1} \L)~, \qquad
\ep {d w \over d t} = \ep \R {d \over d t}(\R^{-1} \w) + \ep^2 \FF(\J) \w~, \label{derp} \feq
\beq \ep \overline{p}(\J) = \R {d \over d t}(\R^{-1} \K)~ \label{deerrpp} \feq
\beq {d s \over d t} = \R {d \over d t}\left(\R^{-1} s \right)
+ \ep \R {d \over d t}\left(\R^{-1} \FF(\J) \v \right)
- \ep^2 (\MM(\J) \v + \FF(\J) \q )~, \label{ders} \feq
\textsl{(note that the right hand sides of Eq.s (\ref{derp}-\ref{ders}) all appear in Eq. \rref{spt4})}. \parn
Eq.s \rref{derp} are mere applications of the general identity \rref{derx} with $\X = \L$
and $\X = w$ (i.e., the function $w(\I,\Te)$), respectively.
Eq. \rref{deerrpp} follows writing \rref{derx} with $\X = \K$, and expressing $d \K/d t =
\ep {d \K/d \tau}$ via Eq. \rref{sistk}~.
The derivation of Eq. \rref{ders} is a bit longer. First of all, from Eq. \rref{derx} with $\X = s$ we infer
\beq {d s \over d t} = \R {d \over d t}(\R^{-1} s) + \ep \FF(\J) s~; \label{derrs} \feq
to continue, we will reexpress $\FF(\J) s$ as $\R \times \!$ a total derivative,
plus terms of the first order in $\ep$. To this purpose,
we write $s$ in terms of $\v$ via Eq. \rref{st}, and then use Eq.
\rref{stp2} with $h=\v$; this gives
\beq s =\omega {\partial \v \over \partial \te}
= {d \v \over d t}  - \ep \Big({\partial \v \over \partial I} f
+ {\partial \v \over \partial \te} g \Big) = {d \v \over d t}  - \ep \q~,
\feq
the last passage following from the definition \rref{dq} of $\q$.
This implies
\beq  \FF(\J) s =
\FF(\J) {d \v \over d t} - \ep \FF(\J)  \q =
{d \over d t} \left(\FF(\J)  \v \right) -  {d\over dt} \left( \FF(\J)\right)  \v
- \ep \,\FF(\J) \q ~.\label{dins} \feq
On the other hand, Eq. \rref{derx} with $\X = \FF(\J) \, \v$ and Eq. \rref{av} give,
respectively,
\beq {d \over d t} \left(\FF(\J) \, \v\right) = \R {d \over d t}\left(\R^{-1}\FF(\J) \, \v\right) +
\ep \left(\FF(\J)\right)^2 \v~; \label{lav0} \feq
\beq {d \over d t} \left({\partial \overline{f} \over \partial I}(\J)\right) =
{\partial^2 \overline{f} \over \partial I^2}(\J) \, {d \J \over d t} =
\ep {\partial^2 \overline{f} \over \partial I^2}(\J) \, \overline{f}(\J)~. \label{lav} \feq
Substituting Eq.s (\ref{lav0}-\ref{lav})
into \rref{dins}, and recalling the definition \rref{em} of $\MM$, we finally get
\beq \ep \FF(\J) s = \ep \R {d \over d t} \left( \R^{-1} \FF(\J) \, \v \right) -
\ep^2 \left(\MM(\J) \, \v + \FF(\J) \, \q \right)~;
\feq
inserting this result into Eq. \rref{derrs}, we obtain the desired relation \rref{ders}. \parn
\textsl{Step 7. One has}
\beq {d \over d t} \,(\R^{-1} \L) =
{d \over d t} \,(\R^{-1}(s - \K )) - \ep {d \over d t}
(\,\R^{-1}(\w - \FF(\J) \v) \,)
+ \label{spt8} \feq
$$  + \ep^2 \,\R^{-1} (\,\u -\FF(\J) (\w +\q) - \MM(\J) \v - \GG(\J, \ep  \L) \L +
{1 \over 2} \HH(\J, \ep \L) \L^2\,)~. $$
To prove this, we return to \rref{spt4} and reexpress
$d \L/ d t - \ep \FF(\J) \L$, $d s /d t$, $\ep dw/d t$, $\ep \overline{p}(\J)$  via Eq.s
(\ref{derp}-\ref{deerrpp}). Multiplying
both sides by $\R^{-1}$, we obtain Eq. \rref{spt8}. \parn
\textsl{Step 8. Conclusion of the proof.}
We integrate Eq. \rref{spt8} from $0$ to $t$, explicitating the dependence
of all objects on $\I,\Te,\J$, $t$ and taking into account the initial
conditions for $\I$, $\Te$, $\J$, $\R$, $\K$, as well as the relations $\L(0)=0$, $\v(I,\te_0)=\w(I,\te_0)=0$.
This gives an expression for $\R^{-1}(\ep t) \, \L(t)$: multiplying by
$\R(\ep t)$, we get the thesis \rref{inseq}.
\section{Appendix. Proof of Lemma \ref{lemxy}.}
\label{alem}
As anticipated, we are inspired by the proof of a
similar statement in \cite{Mitr} (see Chapter XII, $\S$ 23, Theorem 1);
therefore we merely sketch the argument. Let us define
\beq \Ti := \{ \ti \in (0,T)~|~\la(t) < \mi(t) ~~\mbox{for all $t \in [0,\ti)$}~ \}~,
\qquad T_1 := \sup \Ti~. \label{cle} \feq
(Note that \rref{esi} and \rref{ero} give
$\mi(0) > \xx(0,\mi(0)) \geqs 0 = \la(0)$; so, by continuity, $\Ti$ is nonempty). In the sequel
we will assume $T_1 < T$, and infer a contradiction. \parn
From \rref{cle}, it is clear that
$\la(\Tim) \leqs \mi(\Tim)$. We cannot have $\la(\Tim) < \mi(\Tim)$ since
this, by continuity, would be against the definition of $T_1$; thus
\beq \la(\Tim) = \mi(\Tim)~. \label{thente} \feq
On the other hand, the assumptions of the Lemma and \rref{thente} imply
$$ \la(\Tim) \leqs_{(1)} \xx(\Tim, \la(\Tim)) +
\int_{0}^{\Tim} d t' \yy(\Tim, t', \la(t')) =_{(2)} \xx(\Tim, \mi(\Tim)) +
\int_{0}^{\Tim} d t' \yy(\Tim, t', \la(t')) \leqs $$
\beq \leqs_{(3)} \xx(\Tim, \mi(\Tim)) +
\int_{0}^{\Tim} d t' \yy(\Tim, t', \mi(t')) <_{(4)} \mi(\Tim)~, \feq
which gives again a contradiction. (For better clarity: the relations $(1) (2) (3) (4)$
follow, respectively, from \rref{ero}, \rref{thente}, the monotonicity of $\eta$ and \rref{esi}).
\section{Appendix. Proof of Proposition \ref{proprinc}.}
\label{apprinc}
We begin with a Lemma; this holds under the same assumptions written at the beginning
of paragraph 2E, before stating Proposition \ref{mainprop}.
\begin{prop}
\label{cored}
\textbf{Lemma.} Assume that there is a family of functions
$\en_{\delta} \in C([0,U_{\delta}),(0,+\infty))$, labelled by a parameter $\delta \in (0,\delta_{*}]$,
such that the following holds: \parn
i) $U_{\delta} \vain U$ for $\delta \vain 0^{+}$; \parn
ii) for all $\delta \in (0,\delta_*]$ and $\tau \in [0,U_{\delta})$, it is
\beq \en_{\delta}(\tau) < \ro(\tau)/\ep~, \feq
\beq \en_{\delta}(\tau) = \delta +
\alpha(\tau,\ep \en_{\delta}(\tau))
+ \ep | \R(\tau) | \int_{0}^{\tau} d \tau' \, | \R^{-1}(\tau') | \, \gamma(\tau',
\ep \en_{\delta}(\tau'), \en_{\delta}(\tau'))~;  \label{edecont} \feq
iii) for each fixed $\tau \in [0,U)$, the limit $\en(\tau) := \lim_{\delta \vain 0^{+}} \en_{\delta}(\tau) $
exists in $[0,+\infty)$ (note that $\tau \in [0,U_{\delta})$ for sufficiently small $\delta$, due
to i)). \parn
Then the solution $(\I,\Te)$ of the perturbed system exists on
$[0,U/\ep)$ and
\beq | \L(t) | \leqs \en(\ep t) \qquad \mbox{for all $t \in [0,U/\ep)$.} \feq
\end{prop}
\textbf{Proof.} Of course, ii) implies
\beq \en_{\delta}(\tau) >
\alpha(\tau,\ep \en_{\delta} (\tau))
+ \ep | \R(\tau) | \int_{0}^{\tau} d \tau' \,
| \R^{-1}(\tau') | \,  \gamma(\tau', \ep \en_{\delta}(\tau'), \en_{\delta}(\tau'))
\feq
for all $\delta \in (0,\delta_{*}]$ and $\tau \in [0,U_{\delta})$. Therefore, Proposition
\ref{mainprop} can be applied to the function $\en_{\delta}$ on the interval $[0,U_{\delta})$; this implies that
$(\I,\Te)$ exists on $[0,U_{\delta}/\ep)$, and
\beq | \L(t) | < \en_{\delta}(\ep t) \qquad \mbox{for all $t \in [0,U_{\delta}/\ep)$.} \feq
Now, sending $\delta$ to zero and using iii) we easily obtain the thesis. \fine
We now pass to Proposition \ref{proprinc}. So, we have the assumptions at the beginning of paragraph 2E,
strengthened by the smoothness requirements \rref{assug} for $a,b,c,d,e$~.
\vskip 0.1cm\noindent
\textbf{Proof of Proposition \ref{proprinc}.} For the sake of brevity, we put
\beq \aa : \Si \vain \reali, \qquad \ell \mapsto \aa(\ell) :=
\alpha(0, \ep \ell)  \label{aaa} \feq
and extend this definition to any
$\delta \geqs 0$ setting
\beq \aad : \Si \vain \reali~, \qquad \ell \mapsto \aad(\ell) := \aa(\ell) + \delta~. \label{aade} \feq
We proceed in several steps.\parn
\textsl{Step 1. For each $\delta \geqs 0$, $\aad$ is a contractive map.}
In fact, for all $\ell, \ell' \in \Si$, we have
\beq | \aad(\ell) - \aad(\ell') | =
\ep \left|{\partial \alpha \over \partial r}(0, \ep \ell) \right| |\ell - \ell'|  \leqs
\ep M | \ell - \ell' | ~. \label{eppl} \feq
But $\ep M < 1$ by the first inequality \rref{hi}, so the thesis is proved. \parn
\textsl{Step 2. There is $\delta_{*} > 0$ such that,
for all $\delta \in [0, \delta_{*}]$, $\aad$ sends $\Si$ into itself.}
In fact, for any $\delta \geqs 0$ and $\ell \in \Si$,
$$ | \aad(\ell) - \ellu | = | \aa(\ell) + \delta - \ellu | \leqs | \aa(\ell) - \aa(\ellu) | + | \aa(\ellu) - \ellu |
+ \delta \leqs $$
\beq \leqs \ep M | \ell - \ellu | + | \aa(\ellu) - \ellu | + \delta \leqs \ep M \mm + | \aa(\ellu) - \ellu | + \delta~,
\label{ea} \feq
where the second inequality follows from Eq. \rref{eppl} with $\delta=0$.
Now, let us define
\beq \delta_{*} := (1 - \ep M) \mm - | \aa(\ellu) - \ellu |~, \label{eb} \feq
and note that $\delta_{*} > 0$ by \rref{hip}. For $\delta \in [0,\delta_{*}]$ and $\ell \in \Si$, Eq.s
\rref{ea}, \rref{eb} imply $|\aad(\ell) - \ellu| \leqs \mm$, i.e., $\aad(\ell) \in \Si$. \parn
\textsl{Step 3. For all $\delta \in [0, \delta_{*}]$, the map $\aad$ has a unique
fixed point $\ell_{\delta} \in \Si$, which depends continuously on $\delta$}. Existence
and uniqueness of the fixed point follows from the Banach theorem on contractions;
to prove continuity we note that, for all $\delta, \delta' \in [0, \delta_{*}]$,
\beq | \ell_{\delta} - \ell_{\delta'}| = | \aad(\ell_{\delta}) - \aadp(\ell_{\delta'}) | =
| \aa(\ell_{\delta}) + \delta - \aa(\ell_{\delta'}) - \delta' | \leqs \feq
$$ \leqs | \aa(\ell_{\delta}) - \aa(\ell_{\delta'}) | + | \delta - \delta' | \leqs \ep M | \ell_{\delta} - \ell_{\delta'} |
+ | \delta - \delta' |~, $$
the last inequality depending on \rref{eppl} with $\delta=0$. This implies
\beq | \ell_{\delta} - \ell_{\delta'}| \leqs {| \delta - \delta' | \over 1 - \ep M}~; \feq
so the map $\delta \mapsto \ell_{\delta}$ is Lipschitz, and a fortiori continuous. \parn
\textsl{Step 4. Proving the thesis  of i).} This follows from Step 3, with $\delta=0$. \parn
\textsl{Step 5. Proving the thesis of ii)}. For any $\delta \in [0, \delta_{*}]$, let $\ell_{\delta}$ be as in
Step 3. From the standard continuity theorems
for the solutions of a parameter-dependent Cauchy problem, we know that
there is a family $(U_{\delta}, \em_{\delta}, \en_{\delta})_{\delta \in (0,\delta_{*}]}$
with the following properties a) b):  \parn
a) for all $\delta \in (0,\delta_{*}]$, it is
$\em_{\delta}, \en_{\delta} \in C^1([0,U_{\delta}),\reali)$; furthermore, these functions
fulfil the equations
\beq {d \em_\delta \over d \tau} =
| \R^{-1} | \gamma(\cdot, \ep \en_\delta, \en_\delta)~, \qquad
\em_\delta(0) = 0~, \label{tred} \feq
$$ {d \en_\delta \over d \tau} =
\Big(1 - \ep {\partial \alpha \over \partial r}\, (\cdot, \ep \en_\delta )\Big)^{-1}
\left({\partial \alpha \over \partial \tau}\,(\cdot , \ep \en_\delta )
+ \ep | \R | | \R^{-1} | \, \gamma(\cdot , \ep \en_\delta , \en_\delta ) +
\ep | \R |^{-1}~\big(\R \sca {d \R \over d \tau}
\big) \em_\delta \right), $$
\beq \qquad \en_\delta (0) = \ell_\delta \label{quattrod} \feq
with the domain conditions
\beq 0 < \en_{\delta} < \ro/\ep~, \qquad
{\partial \alpha \over \partial r}(\cdot, \ep \en_\delta)  < 1/\ep  ~.\label{dued} \feq
b) One has
\beq U_{\delta} \vain_{\delta \vain 0^{+}} U, ~~\en_\delta(t)
\vain_{\delta \vain 0^{+}} \en(t),~~ \em_\delta(t) \vain_{\delta \vain 0^{+}} \em(t)~~
\mbox{for all $t \in [0, U)$}, \label{du} \feq
where $\em$, $\en$ are as stated in ii). \parn
Let us consider the pair $\em_{\delta}, \en_{\delta}$ for any $\delta \in (0, \delta_{*}]$. Then, integrating
\rref{tred},
\beq \em_{\delta}(\tau) = \int_{0}^{\tau} d \tau' \, | \R^{-1}(\tau')
\, | \gamma(\tau', \ep \en_\delta(\tau'), \en_\delta(\tau')) \qquad \mbox{for $\tau \in [0, U_{\delta})$.}
\label{lass} \feq
Furthermore, from Eq. \rref{quattrod} we infer
$$ 0 = \Big(1 - \ep {\partial \alpha \over \partial r}(\cdot,
\ep \en_\delta)\Big) {d \en_\delta \over d \tau} -
\left({\partial \alpha \over \partial \tau}\,(\cdot , \ep \en_\delta )
+ \ep | \R | | \R^{-1} | \, \gamma(\cdot , \ep \en_\delta , \en_\delta ) +
\ep | \R |^{-1}~\big(\R \sca {d \R \over d \tau}
\big)~\em_\delta \right) = $$
\beq = \Big(1 - \ep {\partial \alpha \over \partial r}(\cdot,
\ep \en_\delta)\Big) {d \en_\delta \over d \tau} -
\left({\partial \alpha \over \partial \tau}\,(\cdot , \ep \en_\delta )
+ \ep | \R | {d \em_{\delta} \over d \tau} +
\ep {d | \R | \over d \tau} \em_\delta \right)~; \label{quattrodd} \feq
the last passage depends on Eq. \rref{tred} for $\em_{\delta}$, and from the identity
$d | \R |/ d \tau = d \sqrt{\R \sca \R} /d \tau = | \R |^{-1} (\R \sca d \R/d \tau)$.
The result \rref{quattrodd} can be rephrased as
\beq 0 = {d \over d \tau} ( \en_{\delta} - \alpha(\cdot, \ep \en_{\delta}) - \ep | \R |
\em_{\delta} )~; \label{daint} \feq
the constant value of the above function can be computed setting $\tau=0$, and is
\beq \en_{\delta}(0) - \alpha(0, \ep \en_{\delta}(0)) =
\ell_{\delta} - \alpha(0, \ep \ell_\delta) = \ell_{\delta} - \aa(\ell_{\delta}) = \delta \feq
(recall the initial condition in \rref{quattrod}, Eq.s \rref{aaa} \rref{aade} and Step 3, giving
$\ell_\delta = \aad(\ell_\delta) = \aa(\ell_\delta) + \delta$). Therefore,
\beq \en_{\delta}(\tau) - \alpha(\tau, \ep \en_{\delta}(\tau)) - \ep | \R(\tau) |
\, \em_{\delta}(\tau) = \delta \qquad \mbox{for $\tau \in [0,U_{\delta})$}~. \label{las} \feq
From Eq.s \rref{las} and \rref{lass}, we see that $\en_\delta$ fulfils Eq. \rref{edecont} of
Lemma \ref{cored}. Due to Eq. \rref{du}
on the limit for $\delta \vain 0^{+}$, from Lemma \ref{cored} we finally obtain the thesis. \fine
\section{Appendix. The functions $\boma{a}$ of the examples.}
\label{apol}
\textbf{Example 1.}
One must determine a function fulfilling Eq. \rref{fa}
for $\tau \in [0,U)$, $\delta J \in (-\J(\tau), \J(\tau))$ and $\te \in \bT$. Neither $\K$ nor $\R$
(nor the initial datum)
play a significant role in this computation, since $\K=0$,
$\ss(I_0,\te_0)=0$ and
$\R(\tau)$ appears in Eq. \rref{fa} as a multiplier for the second of these vanishing terms.
In conclusion, to obtain $a$ we can simply
bind $|s(\J(\tau) + \delta J, \te)|$ in terms of $\J(\tau)$ and $r := |\delta J|$. \parn
Consider any point  $I \in \Lambda$; of course,
\beq \max_{\te \in \bT} | s(I,\te) | = \left(\max_{\te \in \bT} s^2(I,\te) \right)^{1/2}~. \feq
Derivating with respect to $\te$, one finds that the maximum of $s^2$ is attained
for $\cos^2 \te = {1/2} + (1 - \sqrt{1 + 2 I^2})/(4 I)$; by elementary computations,
this gives
\beq \max_{\te \in \bT} | s(I,\te) | = \mbox{a}(I)~, \feq
where $\mbox{a}$ is the $C^{\infty}$, strictly increasing function given by
\beq \mbox{a} : (0, + \infty) \vain (0,+\infty)~, I \mapsto \mbox{a}(I) :
= {1 \over 8} \big(-2 + 10 I^2 + I^4 + 2(1 + 2 I^2)^{3/2}\big)^{1/2}~. \feq
Let $\tau \in [0,U)$,  $\delta J \in (-\J(\tau), \J(\tau))$,  $\te \in \bT$ and $r := | \delta J |$. Then,
\beq | s(\J(\tau) + \delta J,\te) | \leqs \mbox{a}(\J(\tau) + \delta J)
\leqs \mbox{a}(\J(\tau) + r)~; \label{resimp} \feq
the last term above is just the function $a(\tau,r)$ of Table 1. \parn
\textbf{Example 2.} We refer again to  Eq. \rref{fa};
as in the previous example,
$\R(\tau)$ plays no role, since it appears in Eq. \rref{fb} as a multiplier for the term $s(I_0, \te_0) =0$.
A simple computation gives
\beq | s(\J(\tau) + \delta J, \te) - \K(\tau) | = | - {\cc \over 2} (\J(\tau) + \delta J) \sin(2 \te) - \K(\tau) |
\leqs \feq
$$ \leqs {1 \over 2} (\J(\tau) + | \delta J|) + | \K(\tau) |
= {1 \over 2} (\J(\tau) + | \delta J|) - \K(\tau)~; $$
this means that Eq. \rref{fa} is fulfilled by the function $a$ in the Table 2. \parn
\textbf{Example 3.} Again, $\R$ and $\K$ play no role in the analysis of Eq. \rref{fa}
and it suffices to bind
$|s(\J(\tau) + \delta J, \te)|$ in terms of $r := |\delta J|$. Clearly,
\beq |s(\J(\tau) + \delta J, \te)| \leqs {1 \over | \J(\tau) + \delta J |} \leqs
{1 \over \J(\tau) - r}~; \feq
therefore, Eq. \rref{fa} is fulfilled by the function of Table 3. \parn
\textbf{Example 4.}
In the left hand side of Eq. \rref{fa}, the terms $\K(\tau)$ and $s(I_0, \te_0)$ are zero; so, to find $a$ we must bind
$|s(\J(\tau) + \delta J, \te)|$ in terms of $r := |\delta J|$. Maximization with
respect to $\te$ can be done analytically; as a final result, Eq.
\rref{fa} is fulfilled by the function $a$ in Table 4. \parn
In each example, the function $a$ determined as above gives an accurate bound on the left hand side of Eq. \rref{fa}.
\section{Appendix. The functions $\boma{b, c}$ of the examples.}
\label{apoll}
i) In comparison with $a$, the functions $b,c,d,e$ in Eq.s
(\ref{fb}--\ref{fe}) can be constructed using rougher majorizations, see the comments in
paragraph 3A. Here we fix the attention on $b$ and $c$, since the functions
$d, e$ of the examples are obtained trivially. \parn
ii) In all the examples, to find $b$ and $c$ we must essentially derive a majorant
for an expressions of the form
$h(\J(\tau), \delta J, \te)$, where $h(J,\delta J, \te)$ is a trigonometric polynomial in $\te$,
whose coefficients are polynomials in $J$ and $\delta J$.
The majorant should depend only on $\J(\tau)$ and $| \delta J|$;
so, the problem is reduced to finding a function $k$ such that
\beq h(J, \delta J, \te) \leqs k(J, r) \qquad
\mbox{for $\te \in \bT$
and $r := |\delta J|$}~. \label{uh} \feq
Let us exemplify this situation in the construction of $b$; computations for $c$
are quite similar. \parn
\textbf{Example 1.} To find  $b$ we can
bind $\big(\w(\J(\tau)+\delta J, \te) - \FF(\J(\tau))
\, \v(\J(\tau) + \delta J, \te) \big)^2$,
which has the form $h(\J(\tau), \delta J, \te)$ with $h$ a polynomial as above;
the square root of the majorant $k(\J(\tau),r)$ is $b(\tau,r)$. \parn
\textbf{Example 2.} This computation is very similar to the one for Example 1. \parn
\textbf{Example 3.} The left hand side of Eq. \rref{fb} is
\beq  | \w(\J(\tau)+\delta J, \te) - \FF(\J(\tau)) \v(\J(\tau) + \delta J, \te) | =  \feq
$$ = {3 - 4 \cos \te + \cos(2 \te) \over 4 (\J(\tau)+ \delta J)^3} \leqs
{2 \over (\J(\tau) + \delta J)^3} \leqs \left. {2 \over (\J(\tau) - r)^3} \right|_{r = | \delta J |}~. $$
\textbf{Example 4.} In this case,
\beq | \w(\J(\tau) + \delta J, \te) - \FF(\J(\tau)) \,
\v(\J(\tau) + \delta J, \te) |^2 = {h(\J(\tau), \delta J, \te) \over (\J^1(\tau) +
\delta J^1)^4 (\J^2(\tau) + \delta J^2)^4} \feq
with $h$ a polynomial as before. After finding for $h$ a bound of the
form \rref{uh}, we combine it with the obvious relation $(\J^i(\tau) + \delta J^i)^{-4}
\leqs (\J^i(\tau) - r)^{-4}$ for $r := |\delta J|$; the square root of the final majorant is
$b(\tau,r)$. \parn
iii) Up to now, we have not explained how to get elementary bounds of the form \rref{uh} on
a polynomial $h$. Here we illustrate a general procedure (computations to apply it in Examples 1-4
are generally too tedious to be made by hand, but are easily implemented on MATHEMATICA). \parn
iii a) In the expression of $h(J, \delta J, \te)$, if $d=1$
we put $\delta J = r \cos \psi$ with $\psi = 0$ or $\pi$;
if $d=2$, we set $\delta J = (r \cos \psi, r \sin \psi)$ with $\psi \in \bT$. \parn
iii b) Now, $h(J, \delta J, \te)$ has the form of a trigonometric polynomial
in $\te, \psi$ with coefficients depending on $J$. We write this in a canonical form,
reexpressing any term in $\te$ and $\psi$ as a linear combination of sines and cosines
(e.g., $\cos^4 \te \sin(2 \te)^2 = (1/32)(5 + 4 \cos(2 \te) - 4 \cos(4 \te)
- 4 \cos(6 \te) - \cos(8 \te))$. \parn
iii c) As a final step, we bind each summand of $h$ using the relations $|\cos|, |\sin| \leqs 1$.
\vfill \eject \noindent
\textbf{Acknowledgments.} We gratefully acknowledge the anonymous referees for useful suggestions about the
style and organization of the paper. \parn
This work has been partially supported by the GNFM
of Istituto Nazionale di Alta Ma\-te\-ma\-ti\-ca and by MIUR,
Research Project Cofin/2004 "Metodi geometrici nella teoria delle onde non lineari e applicazioni".
\salto

\end{document}